\documentclass[english,nofootinbib]{revtex4-2}
\usepackage[T1]{fontenc}
\usepackage[latin9]{inputenc}
\usepackage{xcolor}
\usepackage{array}
\usepackage{mathtools}
\usepackage{dsfont}
\usepackage{amsmath}
\usepackage{stackrel}
\usepackage{graphicx}

\makeatletter

\providecommand{\tabularnewline}{\\}

\usepackage{babel}

\makeatother

\usepackage{babel}
\begin{document}
\title{Simultaneous quantum identity authentication scheme utilizing entanglement
swapping with secret key preservation }
\author{Arindam Dutta}
\email{arindamsalt@gmail.com}

\email{https://orcid.org/0000-0003-3909-7519}

\author{Anirban Pathak}
\email{anirban.pathak@gmail.com}

\email{https://orcid.org/0000-0003-4195-2588}

\affiliation{Department of Physics and Materials Science \& Engineering, Jaypee
Institute of Information Technology, A 10, Sector 62, Noida, UP-201309,
India}
\begin{abstract}
Unconditional security in quantum key distribution (QKD) relies on
authenticating the identities of users involved in key distribution.
While classical identity authentication schemes were initially utilized
in QKD implementations, concerns regarding their vulnerability have
prompted the exploration of quantum identity authentication (QIA)
protocols. In this study, we introduce a new protocol for QIA, derived
from the concept of controlled secure direct quantum communication.
Our proposed scheme facilitates simultaneous authentication between
two users, Alice and Bob, leveraging Bell states with the assistance
of a third party, Charlie. Through rigorous security analysis, we
demonstrate that the proposed protocol withstands various known attacks,
including impersonation, intercept and resend and impersonated fraudulent
attacks. Additionally, we establish the relevance of the proposed
protocol by comparing it with the existing protocols of similar
type.
\end{abstract}
\maketitle

\section{Introduction}

With the widespread adoption of online banking, e-commerce platforms
and Internet of Things (IoT) devices, end-users have become accustomed
to identity authentication procedures. The increased frequency of
using such applications has underscored the importance of identity
authentication schemes, which are systematic methodologies for verifying
the identity of legitimate users or components/devices. These schemes
are integral to virtually all cryptographic tasks, including securing
communication and computation processes.

Quantum cryptography has notably revolutionized the concept of security
by offering unconditional security, a feature unattainable in the
classical realm. Bennett and Brassard's seminal work in 1984 introduced
an unconditionally secure protocol for quantum key distribution (QKD)
\cite{BB84}. Unlike classical protocols, which rely on computational
complexity for security, the security of the BB84 protocol stems from
the principles of quantum mechanics. Subsequent QKD protocols (\cite{E91,B92,WSL+21,SZ22,DP+23,DMB+24}
and others \cite{LCQ12,CLY+20,JJL+13,HAL+20,WYH+22} enable secure
key distribution between a sender (Alice) and a receiver (Bob). However,
before initiating any QKD protocol, Alice and Bob must authenticate
each other to prevent impersonation by an eavesdropper (Eve). Therefore,
identity authentication is pivotal for implementing QKD and other
quantum cryptographic tasks. Initially, classical authentication schemes
like the Wegman--Carter scheme \cite{WC81} were mentioned as the
scheme to be used in implementing the authentication step in a QKD
protocol. Even in Bennett and Brassard's pioneering work \cite{BB84},
it was mentioned that identity authentication will be performed by
using Wegman--Carter scheme. However, classical authentication schemes
lack unconditional security. Presently, commercially available QKD
products and most laboratory-level implementations of quantum cryptographic
tasks rely on classical authentication schemes, including post-quantum
schemes. Consequently, protocols for QKD and other quantum cryptographic
tasks are not truly unconditionally secure or fully quantum unless
unconditionally secure schemes for quantum identity authentication
(QIA) are developed. Motivated by this necessity, researchers have
proposed numerous QIA schemes \cite{CS_1995,LB_2004,WZT_2006,ZZZX_2006,ZCSL_2020,KHHYHM_2018,CXZY_2014,T.Mihara_2002,DP22,DP23,JWC+23,LZZ+22},
leveraging quantum resources to achieve desired security levels. These
protocols typically initiate with a pre-shared small key used for
authentication. Some efforts have explored the possibility of designing
schemes for device-independent QIA \cite{FG21} and QIA that use homomorphic
encryption with qubit rotation \cite{CWJ+23}. 

The first QIA scheme was introduced by Cr{\'e}peau et al. \cite{CS_1995}
in 1995, utilizing oblivious transfer (OT) as a cryptographic primitive.
However, subsequent work by Lo and Chau \cite{LC_1997} demonstrated
that quantum OT cannot achieve unconditional security in a two-party
scenario. Consequently, Cr{\'e}peau et al.'s QIA scheme lacked unconditional
security. It is noteworthy that OT is not the sole cryptographic task
applicable to QIA scheme design. Following Cr{\'e}peau et al.'s work,
a series of QIA protocols emerged \cite{LB_2004,WZT_2006,ZZZX_2006,ZCSL_2020,KHHYHM_2018,CXZY_2014,T.Mihara_2002,ZW_1998,DHHM_1999,DP22}.
The schemes for QIA are built upon protocols designed for various
cryptographic tasks, serving as the foundation for their development.
For a comprehensive understanding of QIA and the adaptation of diverse
cryptographic tasks to formulate QIA schemes, refer to the review
provided in Ref. \cite{DP22}. Some of these tasks were adapted from
protocols for secure direct quantum communication, which involve transmitting
messages using quantum resources without key generation \cite{LL02,LM_05,P13}.
Before delving into a summary of existing QIA schemes derived from
modified secure direct communication schemes, it is pertinent to distinguish
between two types of secure quantum direct communication: quantum
secure direct communication (QSDC) and deterministic secure quantum
communication (DSQC). Numerous QSDC and DSQC schemes exist \cite{LL02,STP20,STP17,YSP14,BP12,DLL03,DP24,PMS+23},
with many early ones being entangled state-based \cite{LL02,DLL03,BF02}.
For instance, Zhang et al. \cite{ZZZX_2006} adapted the entangled
state-based ping-pong protocol \cite{BF02} for QSDC to devise a QIA
scheme. Moreover, Yuan et al. \cite{YLP+14} proposed a QIA protocol
based on the LM05 protocol \cite{LM_05}, which is the single photon
analog of the ping-pong protocol. Other authors have also modified
DSQC and QSDC protocols to create QIA schemes \cite{LB_2004,ZLG_2000,LC_2007}.
Additionally, novel schemes have emerged for controlled DSQC (CDSQC)
\cite{STP17,P15}, where a semi-honest Alice securely communicates
with Bob using quantum resources only under the supervision of a controller,
Charlie. The potential of adapting CDSQC schemes for QIA remains relatively
unexplored. Taking into account the significance of identity authentication,
we aim to propose a new QIA protocol with simultaneous authentication
of legitimate parties, inspired by CDSQC principles. In the following
sections, we will introduce a Bell state-based protocol for QIA, drawing
from CDSQC concepts, and demonstrate its resilience against common
attacks. Additionally, we will analyze the impact
of collective noise on the proposed scheme.

The remainder of this letter is structured as follows: Section \ref{sec:II}
introduces the proposed QIA protocol that uses Bell states. Section
\ref{sec:III} critically analyzes the protocol's security, demonstrating
its robustness against multiple types of attacks. Section
\ref{sec:IV} focuses on collective noise analysis for the new QIA
scheme. Section \ref{sec:V} compares our protocol with the existing
ones from the same family. Finally, Section \ref{sec:VI} concludes
the letter.

\section{New quantum identity authentication protocol\label{sec:II}}

In this section, we aim to introduce a new framework for QIA, akin
to schemes inspired by QSDC. As it utilizes Bell states, our protocol
can be construed as an entangled state-driven approach to QIA. Within
this framework, two legitimate parties mutually authenticate one another
with the assistance of an untrusted third party, employing unitary
operations. Prior to delving into the specifics of our protocol, it
is important to provide a succinct overview of the underlying conceptual
framework driving its design.

\subsection{Principal concept\label{subsec:Principal-concept}}

This protocol relies on the principles of Bell states with entanglement
swapping and the application of Pauli operations. The corresponding
connections between the Bell state and the pre-shared authentication
key can be delineated as follows:

\begin{equation}
\begin{array}{lcl}
00:|\phi^{+}\rangle & = & \frac{1}{\sqrt{2}}\left(\left|00\right\rangle +\left|11\right\rangle \right),\\
01:|\phi^{-}\rangle & = & \frac{1}{\sqrt{2}}\left(\left|00\right\rangle -\left|11\right\rangle \right),\\
10:|\psi^{+}\rangle & = & \frac{1}{\sqrt{2}}\left(\left|01\right\rangle +\left|10\right\rangle \right),\\
11:|\psi^{-}\rangle & = & \frac{1}{\sqrt{2}}\left(\left|01\right\rangle -\left|10\right\rangle \right).
\end{array}\label{eq:Bell-State with Pre-Shared_Key}
\end{equation}
The Eqs. (\ref{eq:Bell-State with Pre-Shared_Key}) and (\ref{eq:Pauli-Operation with Pre-Shared_Key})
describe the relationship between the pre-shared key, and the Bell
state prepared by Alice and Bob, as well as the Pauli operations they
perform. The association between the respective Pauli operations and
the pre-shared authentication key can be stated as follows:

\begin{equation}
\begin{array}{lcl}
00:\mathds{1}_{2} & = & \left|0\right\rangle \left\langle 0\right|+\left|1\right\rangle \left\langle 1\right|,\\
01:\sigma_{x} & = & \left|0\right\rangle \left\langle 1\right|+\left|1\right\rangle \left\langle 0\right|,\\
10:i\sigma_{y} & = & \left|0\right\rangle \left\langle 1\right|-\left|1\right\rangle \left\langle 0\right|,\\
11:\sigma_{z} & = & \left|0\right\rangle \left\langle 0\right|-\left|1\right\rangle \left\langle 1\right|.
\end{array}\label{eq:Pauli-Operation with Pre-Shared_Key}
\end{equation}
Let us provide a concise overview of the protocol described. Alice
and Bob possess a pre-shared authentication key sequence consisting
of $n+1$ pairs of bits. Each key pair represents two bits of information.
Initially, let us consider a specific example where we denote the
first two secret keys of the sequence as $11$ and $00$. Consider
that Alice selects the second key ($00$), generally referred to as
the $\left(m+1\right)^{{\rm th}}$ key in the entire sequence, while
Bob selects the XOR of the second key and the first key $\left(00\oplus11=11\right)$,
which can be understood as the XOR of the $\left(m+1\right)^{{\rm th}}$
key and the $m^{{\rm th}}$ key, where $m=1,2,\cdots,n.$ Subsequently,
they generate Bell states based on their selected keys using the Eq.
(\ref{eq:Bell-State with Pre-Shared_Key}). In this scenario, Alice
and Bob prepare the states $\left|\phi^{+}\right\rangle _{12}$ and
$\left|\psi^{-}\right\rangle _{34}$, where the subscripts $1,2$
and $3,4$ represent the particles of Alice and Bob, respectively.
Similarly, they generate a series of Bell states using the pre-shared
authentication key sequence. Alice retains the sequence of particle
$1$ of the Bell states and transmits the sequence associated with
particle $2$ of the Bell states to an untrusted third party, Charlie.
Charlie applies a permutation operator $\Pi_{n}$ \cite{YSP14,TP15,SPS12,P15}
to the sequence of particle $2$ while keeping the original sequence
confidential. Following the permutation, Charlie forwards the sequence
to Bob. Bob independently prepares two sequences associated with particle
$3$ and particle $4$. Bob sends the sequence corresponding to particle
$4$ directly to Alice. Consequently, Alice and Bob possess the sequences
containing particles $1,4$ and $2,3$, respectively. With the specific
two secret keys in place, the composite system is characterized by,

\[
\begin{array}{lcl}
\left|\phi^{+}\right\rangle _{12}\left|\psi^{-}\right\rangle _{34} & = & \frac{1}{2}\left(\left|\psi^{+}\right\rangle _{14}\left|\phi^{-}\right\rangle _{23}+\left|\psi^{-}\right\rangle _{14}\left|\phi^{+}\right\rangle _{23}-\left|\phi^{+}\right\rangle _{14}\left|\psi^{-}\right\rangle _{23}-\left|\phi^{-}\right\rangle _{14}\left|\psi^{+}\right\rangle _{23}\right).\end{array}
\]
Alice (Bob) applies a Pauli operation $\sigma_{z}$ on qubit $1$
(qubit $3$) in accordance with the first key, here considered as
$11.$ This\footnote{For a long pre-shared key sequence, the Pauli operation will be determined
by the $m^{{\rm th}}$ key.} is done utilizing the map described in Eq. (\ref{eq:Pauli-Operation with Pre-Shared_Key}).
Subsequently, Alice and Bob publicly announce the completion of their
operations on the designated particles. Charlie then publicly announces
the permutation operation ($\Pi_{n}$). Bob performs the inverse permutation
operation on his sequence of particle 2, thereby restoring the original
sequence. Following this, they conduct a Bell measurement on the particles
currently in their possession (Alice performs Bell measurement on
particles $1$ and $4$, while Bob on particles $2$ and $3$). The
resultant composite system is as follows,

\begin{equation}
\begin{array}{lcl}
\left|\Psi^{11}\right\rangle  & = & \sigma_{z1}\otimes\sigma_{z3}\left(\left|\phi^{+}\right\rangle _{12}\left|\psi^{-}\right\rangle _{34}\right)\\
 & = & \frac{1}{2}\left(\left|\psi^{-}\right\rangle _{14}\left|\phi^{+}\right\rangle _{23}+\left|\psi^{+}\right\rangle _{14}\left|\phi^{-}\right\rangle _{23}-\left|\phi^{-}\right\rangle _{14}\left|\psi^{+}\right\rangle _{23}-\left|\phi^{+}\right\rangle _{14}\left|\psi^{-}\right\rangle _{23}\right).
\end{array}\label{eq:Final composite system=00007B11=00007D}
\end{equation}
Alice (Bob) converts the outcome of her (his) Bell measurement into
classical bits using Eq. (\ref{eq:Bell-State with Pre-Shared_Key}).
The measurement result is not disclosed. In our scenario, Alice has
measurement outcomes, each with a probability of $\frac{1}{4}$, represented
as $\left|\psi^{-}\right\rangle ,$ $\left|\psi^{+}\right\rangle ,$
$\left|\phi^{-}\right\rangle $ and $\left|\phi^{+}\right\rangle $,
with corresponding classical bit values of 11, 10, 01 and 00. The
measurement outcomes of Bob correspond to $\left|\phi^{+}\right\rangle ,$
$\left|\phi^{-}\right\rangle ,$ $\left|\psi^{+}\right\rangle $ and
$\left|\psi^{-}\right\rangle $, with corresponding classical bit
values 00, 01, 10 and 11, respectively. In this context, Alice assumes
the role of the verifier since we are considering two secret keys
in our example. Suppose Bob declares his bit value as 00 through an
unjammable classical communication channel. Then Alice performs an
XOR operation on Bob's announcement and her classical bit value (11)
that corresponds to her measurement outcome. If no errors occur in
the quantum channel due to noise or eavesdropping, the XOR value should
match the first key of the initial two secret key sequences, which
is 11. It is important to note that generally, the XOR value would
correspond to the $m^{{\rm th}}$ key when considering the pre-shared
key sequence. The remaining possible outcomes and their correlations
are detailed in Table \ref{tab:The-possible-measurement} when the
initial pre-shared keys are 11 and 00.

\begin{table}[h]
\begin{centering}
\begin{tabular*}{14cm}{@{\extracolsep{\fill}}cc}
\hline 
Alice and Bob's possible measurement outcomes & Additional modulo 2 \tabularnewline
\hline 
$\left|\psi^{-}\right\rangle _{14}\otimes\left|\phi^{+}\right\rangle _{23}$ & $11\oplus00=11$\tabularnewline
$\left|\psi^{+}\right\rangle _{14}\otimes\left|\phi^{-}\right\rangle _{23}$ & $10\oplus01=11$\tabularnewline
$\left|\phi^{-}\right\rangle _{14}\otimes\left|\psi^{+}\right\rangle _{23}$ & $01\oplus10=11$\tabularnewline
$\left|\phi^{+}\right\rangle _{14}\otimes\left|\psi^{-}\right\rangle _{23}$ & $00\oplus11=11$\tabularnewline
\hline 
\end{tabular*}
\par\end{centering}
\caption{\label{tab:The-possible-measurement}The possible measurement results
of the legitimate parties for our QIA scheme.}

\end{table}

We take 3 more examples to ensure the persistence of our conclusion
across the remaining three possible secret keys in the $m^{{\rm th}}$
position of the pre-shared key sequence, namely \{00\}, \{01\} and
\{10\}, we can depict the final state shared by Alice and Bob. The
final composite systems, considering the secret keys at positions
$m^{{\rm th}}$ and $\left(m+1\right)^{{\rm th}}$ of the entire sequence
to be $\left\{ 00,01\right\} ,$ $\left\{ 01,10\right\} $ and $\left\{ 10,11\right\} $
respectively, are expressed as follows:
\begin{equation}
\begin{array}{lcl}
\left|\Psi^{00}\right\rangle  & = & \mathds{1}_{1}\otimes\mathds{1}_{3}\left(\left|\phi^{-}\right\rangle _{12}\left|\phi^{-}\right\rangle _{34}\right)\\
 & = & \frac{1}{2}\left(\left|\phi^{+}\right\rangle _{14}\left|\phi^{+}\right\rangle _{23}+\left|\phi^{-}\right\rangle _{14}\left|\phi^{-}\right\rangle _{23}-\left|\psi^{+}\right\rangle _{14}\left|\psi^{+}\right\rangle _{23}-\left|\psi^{-}\right\rangle _{14}\left|\psi^{-}\right\rangle _{23}\right)
\end{array},\label{eq:Final composite system=00007B00=00007D}
\end{equation}

\begin{equation}
\begin{array}{lcl}
\left|\Psi^{01}\right\rangle  & = & \sigma_{x1}\otimes\sigma_{x3}\left(\left|\psi^{+}\right\rangle _{12}\left|\psi^{-}\right\rangle _{34}\right)\\
 & = & \frac{1}{2}\left(-\left|\phi^{+}\right\rangle _{14}\left|\phi^{-}\right\rangle _{23}-\left|\phi^{-}\right\rangle _{14}\left|\phi^{+}\right\rangle _{23}+\left|\psi^{+}\right\rangle _{14}\left|\psi^{-}\right\rangle _{23}+\left|\psi^{-}\right\rangle _{14}\left|\psi^{+}\right\rangle _{23}\right)
\end{array},\label{eq:Final composite system=00007B01=00007D}
\end{equation}
and
\begin{equation}
\begin{array}{lcl}
\left|\Psi^{10}\right\rangle  & = & i\sigma_{y1}\otimes i\sigma_{y3}\left(\left|\psi^{-}\right\rangle _{12}\left|\phi^{-}\right\rangle _{34}\right)\\
 & = & \frac{1}{2}\left(\left|\psi^{+}\right\rangle _{14}\left|\phi^{+}\right\rangle _{23}+\left|\psi^{-}\right\rangle _{14}\left|\phi^{-}\right\rangle _{23}+\left|\phi^{+}\right\rangle _{14}\left|\psi^{+}\right\rangle _{23}+\left|\phi^{-}\right\rangle _{14}\left|\psi^{-}\right\rangle _{23}\right)
\end{array},\label{eq:Final composite system=00007B10=00007D}
\end{equation}
respectively. In Eqs. (\ref{eq:Final composite system=00007B11=00007D}
- \ref{eq:Final composite system=00007B10=00007D}), we consider 4
combinations of $m^{{\rm th}}$ and $\left(m+1\right)^{{\rm th}}$
terms. However, there are 12 additional combinations that yield the
same conclusive measurement results which provide authentication for
legitimate parties.

\subsection{Protocol description}

\begin{figure}
\begin{centering}
\includegraphics[scale=0.5]{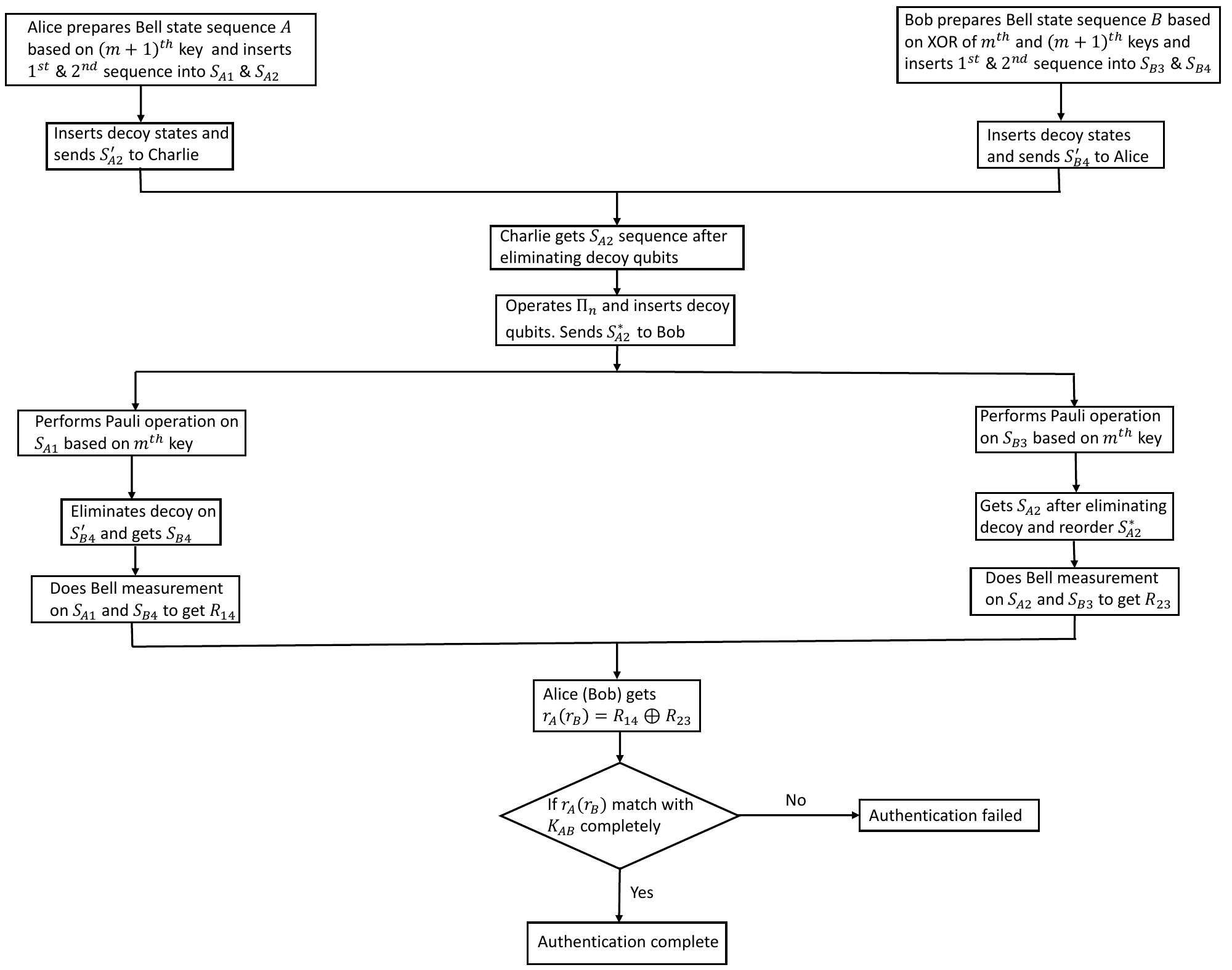}
\par\end{centering}
\caption{\label{fig:Figure_1}Flowchart illustrating the operation of the proposed
QIA protocol.}
\end{figure}

In the scenario considered in this letter, two legitimate parties,
Alice and Bob, seek to authenticate themselves with the help of an
untrusted party Charlie. Within this authentication protocol, Alice
and Bob possess a pre-shared classical secret key denoted as $K_{AB}=\left\{ k_{1}^{1}k_{2}^{1},k_{1}^{2}k_{2}^{2},k_{1}^{3}k_{2}^{3},\cdots,k_{1}^{m}k_{2}^{m},\cdots k_{1}^{n+1}k_{2}^{n+1}\right\} $.
The secret key sequence $K_{AB}$ comprises two bit key information
with uniform distribution, denoted by $k_{1}^{m}k_{2}^{m}$, where
$k_{1}^{m}k_{2}^{m}\in\{00,01,10,11\}$. Before delving into the protocol's
procedural details, it is pertinent to briefly elucidate the utility
of decoy states in establishing secure communication channels between
any two parties \cite{LMC05}. Typically, one party randomly inserts
decoy qubits, ideally equal in number to the original information
sequence, into the information qubit sequence. This enlarged sequence
is then transmitted to the other party. Upon receiving this enlarged
sequence, the receiving party asks for positional and encoding details
of the decoy qubits from the transmitting party via an impervious
public channel. The transmitting party declares this requisite information
publicly, enabling the receiving party to verify it by measuring the
decoy qubits. Should errors exceeding the acceptable threshold be
detected, both parties terminate the entire qubit sequence. This utilization
of decoy states heightens the security of the channel and facilitates
the secure transmission of information between parties. Now, let's
delineate the sequential steps involved in the authentication process:
\begin{description}
\item [{Step~1}] Alice and Bob generate Bell state sequences $A_{12}$
and $B_{34}$ using the $\left(m+1\right)^{{\rm th}}$ key and XOR
operation of the $m^{{\rm th}}$ and $\left(m+1\right)^{{\rm th}}$
keys of the entire sequence $K_{AB}$, respectively, as stipulated
by the procedure outlined in Eq. (\ref{eq:Bell-State with Pre-Shared_Key}),
where $m=1,2,\cdots,n$.\\
\[
\begin{array}{lcl}
A & = & \left\{ |A\rangle_{12}^{1},|A\rangle_{12}^{2},|A\rangle_{12}^{3},\cdots,|A\rangle_{12}^{m},\cdots,|A\rangle_{12}^{n}\right\} ,\\
B & = & \left\{ |B\rangle_{34}^{1},|B\rangle_{34}^{2},|B\rangle_{34}^{3},\cdots,|B\rangle_{34}^{m},\cdots,|B\rangle_{34}^{n}\right\} .
\end{array}
\]
The subscripts $1,2$ and $3,4$ correspond to the particles assigned
to Alice and Bob, respectively. Ideally, these two sets should be
identical.
\item [{Step~2}] Alice and Bob partition the states of their Bell pairs
into two distinct sequences of $n$ particles. In each sequence, the
first particle of every Bell pair constitutes one sequence, while
the second particle of each Bell pair forms the other sequence. Thus,
Alice and Bob possess the sequences denoted as $S_{A1},\,S_{A2}$
and $S_{B3},\,S_{B4}$, respectively.\\
\[
\begin{array}{lcl}
S_{A1} & = & \left\{ s_{1}^{1},s_{1}^{2},s_{1}^{3},\cdots,s_{1}^{m},\cdots,s_{1}^{n}\right\} ,\\
S_{A2} & = & \left\{ s_{2}^{1},s_{2}^{2},s_{2}^{3},\cdots,s_{2}^{m},\cdots,s_{2}^{n}\right\} ,\\
S_{B3} & = & \left\{ s_{3}^{1},s_{3}^{2},s_{3}^{3},\cdots,s_{3}^{m},\cdots,s_{3}^{n}\right\} ,\\
S_{B4} & = & \left\{ s_{4}^{1},s_{4}^{2},s_{4}^{3},\cdots,s_{4}^{m},\cdots,s_{4}^{n}\right\} .
\end{array}
\]
Here, $S_{A1}$ ($S_{A2}$) represents the sequence comprising the
first (second) particles of all Bell states within sequence $A$.
Correspondingly, $S_{B3}$ ($S_{B4}$) denotes the sequence encompassing
the first (second) particles of all Bell states within sequence $B$.
Alice and Bob possess the particle sequences $S_{A1}$ and $S_{B3}$
with themselves, respectively. Additionally, Alice (Bob) procedurally
introduces the decoy particles $D_{A}$ $\left(D_{B}\right)$ into
the sequence $S_{A2}$ ($S_{B4}$) to formulate an augmented sequence
$S_{A2}^{\prime}$ ($S_{B4}^{\prime}$). Alice (Bob) subsequently
transmits the sequence $S_{A2}^{\prime}$ ($S_{B4}^{\prime}$) via
the quantum channel to Charlie (Alice).
\item [{Step~3}] Charlie receives the sequence $S_{A2}^{\prime}$ and
performs security tests utilizing decoy particles. Following the successful
completion of the security tests, Charlie eliminates the decoy particles
$D_{A}$ to retrieve the original sequence $S_{A2}$. Upon obtaining
$S_{A2}$, Charlie applies a permutation operation $\Pi_{n}$ to the
sequence and introduces decoy states to produce the new sequence $S_{A2}^{*}$.
Subsequently, Charlie transmits the new sequence $S_{A2}^{*}$ to
Bob.
\item [{Step~4}] Alice and Bob perform a security test employing decoy
particles. Upon the successful conclusion of the security assessment,
Alice eliminates the decoy states denoted as $D_{B}$ from the enlarged
sequence $S_{B4}^{'}$, reverting it to the initial sequence prepared
by Bob, designated as $S_{B4}$. Subsequently, Charlie publicly discloses
the precise ordering of the sequence $S_{A2}^{*}$ which is a reverse
operation of $\Pi_{n}$ operator. Bob reorganizes the sequence to
restore its original configuration, denoted as $S_{A2}$.
\item [{Step~5}] Alice (Bob) applies Pauli operations to all $m^{{\rm th}}$
particles in sequence $S_{A1}$ ($S_{B3}$) based on the $m^{{\rm th}}$
key of the pre-shared key sequence, utilizing Eq. (\ref{eq:Pauli-Operation with Pre-Shared_Key}).
Subsequently, Alice (Bob) possesses sequences $S_{A1},\,S_{B4}$ ($S_{A2},\,S_{B3}$)
and conducts Bell measurements on the respective particles of these
sequences\footnote{Perform Bell measurement on $m^{{\rm th}}$ particle of sequence $S_{A1}$($S_{A2}$)
and $m^{{\rm th}}$ particle of sequence $S_{B4}$ ($S_{B3}$).}. Alice and Bob record the measurement outcomes of the Bell state
measurement as classical key sequences $R_{14}=\left\{ r_{14}^{1},r_{14}^{2},r_{14}^{3},\cdots,r_{14}^{m},\cdots,r_{14}^{n}\right\} $
and $R_{23}=\left\{ r_{23}^{1},r_{23}^{2},r_{23}^{3},\cdots,r_{23}^{m},\cdots,r_{23}^{n}\right\} $
following Eq. (\ref{eq:Bell-State with Pre-Shared_Key}). 
\item [{Step~6}] Alice publicly discloses the position and values of $\frac{n}{2}$
keys from the sequence $R_{14}$, while Bob also discloses $\frac{n}{2}$
keys from the sequence $R_{23}$ of corresponding remaining position
of the sequence $R_{14}$. Alice (Bob) executes XOR operations between
Bob's (Alice's) announced key and the corresponding keys from her
(his) sequence $R_{14}$ ($R_{23}$), resulting in a new two-bit key
sequence $r_{A}$ ($r_{B}$) of length $n$.
\item [{Step~7}] In an error-free scenario, the elements of $r_{A}$ and
$r_{B}$ should match with the corresponding key elements of the initial
sequence $K_{AB}$\footnote{Here, $n$ elements of sequences of $r_{A}$ and $r_{B}$ should match
with the corresponding $n$ elements of sequence of $K_{AB}$. It
may be noted that $K_{AB}$ has $n+1$ pre-shared keys.}. Upon meeting this criterion, authentication is successfully attained
for both Alice and Bob. This authentication process is executed simultaneously
by the two legitimate parties.\\
The presented QIA protocol is depicted in a flowchart illustrated
in Fig. \ref{fig:Figure_1}.
\end{description}

\section{Evaluation of the security aspects in the proposed protocol\label{sec:III}}

In this section, we aim to evaluate the security of the proposed protocol
against various well-known attacks that can be attempted by an eavesdropper,
commonly referred to as Eve. It is important to mention that insider
attacks (by Alice and Bob) will not be discussed here, as they are
not pertinent within the scope of QIA. Initially, we will focus on
the impersonation attack. In this scenario, we consider the possibility
of an untrusted participant, Charlie, attempting to gain unauthorized
access to confidential information while adhering to all steps of
the protocol. Moreover, an impersonated fraudulent attack entails
an external party, such as Eve, aiming to mimic the legitimate user,
either Alice or Bob, and effectively maneuver through the authentication
procedure. In particular,
among the attacks discussed in this section, the intercept-resend
attack and impersonation-based fraudulent attack specifically target
the quantum channel.

\subsection{Security assessment of the protocol against impersonation attack
by Eve}

Without loss of generality, let us consider the scenario outlined
in Section \ref{subsec:Principal-concept}. Eve, acting as Alice,
selects the state $|\phi^{+}\rangle_{12}$ and transmits the particle
(particle 2) sequence $S_{e2}$ to Charlie\footnote{We adopt the term \textquotedbl sequence\textquotedbl{} for the sake
of generality. However, to elucidate the intricacies of this security
analysis, we employ an illustrative example outlined in Section \ref{subsec:Principal-concept}.}. Charlie adheres to the protocol by applying the permutation operation
and forwards the sequence $S_{e2}^{\prime}$ to Bob. Upon receiving
Charlie's announcement regarding the correct sequence order of $S_{e2}^{\prime}$,
Bob retrieves the original $S_{e2}$. Additionally, Bob sends the
particle sequence $S_{B4}$ directly to Eve. Eve, while masquerading
as Alice, executes all protocol steps as expected, applying the Pauli
operation (here $\sigma_{z1}$) to Particle $1$. Similarly, Bob performs
the Pauli operation (here $\sigma_{z3}$) on his particle 3. Subsequently,
Bob and Eve conduct Bell measurements, and Table (\ref{tab:The-possible-measurement})
illustrates all possible measurement scenarios revealing Eve's presence.

\begin{figure}
\begin{centering}
\includegraphics[scale=0.5]{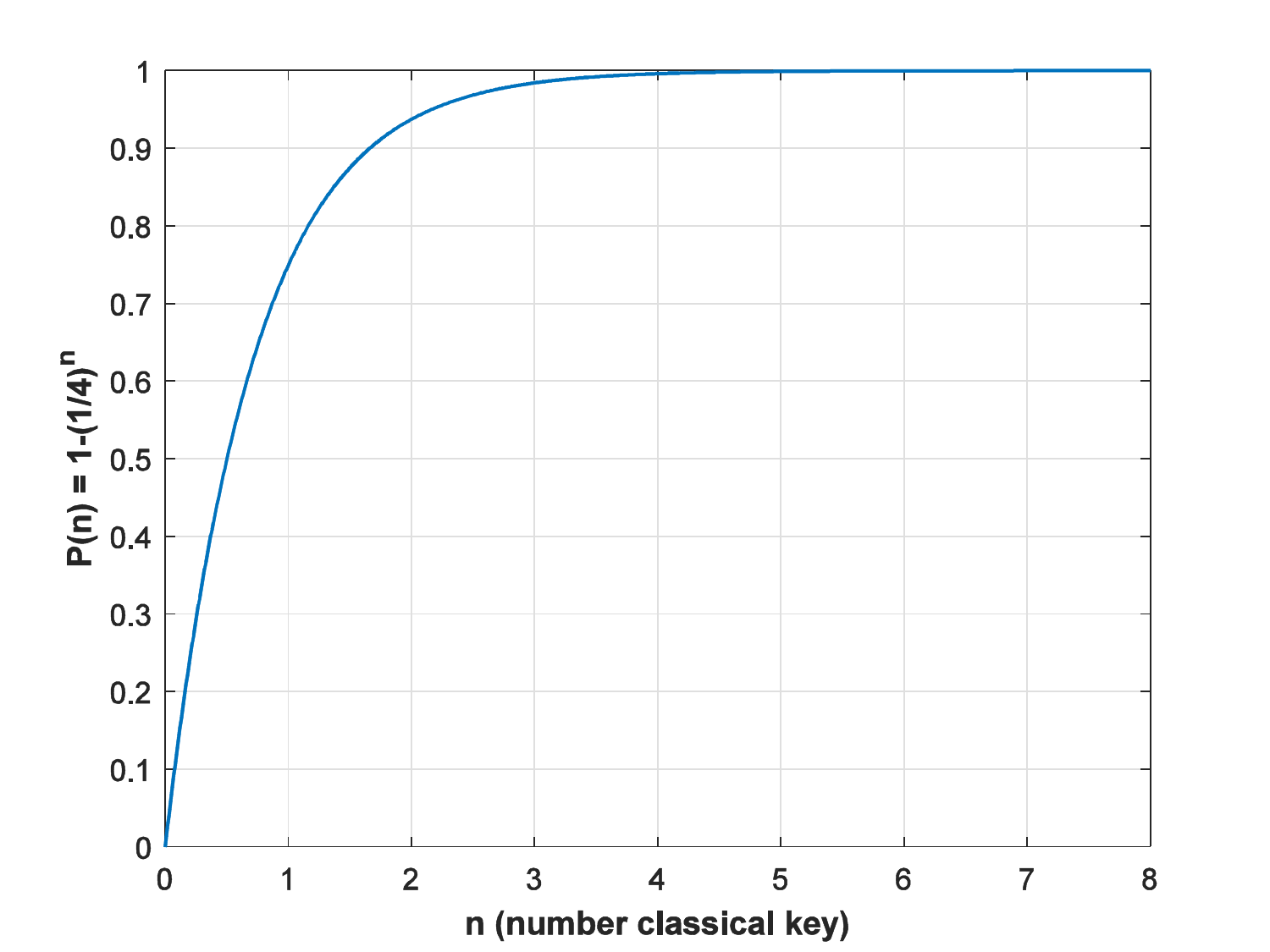}
\par\end{centering}
\caption{\label{fig:Figure_2}The correlation between the probability $P\left(n\right)$
of detecting Eve\textquoteright s presence and the number of classical
keys $n$ used as a pre-shared authentication key.}
\end{figure}

As Eve does not possess knowledge of the correct key pair $K_{AB}$
(e.g., 11 and 00) nor the permutation operation performed by Charlie,
$\Pi_{n}$, her probability of selecting the correct Bell state is
$\frac{1}{4}$. To successfully evade detection, Eve must accurately
guess all $n$ Bell states correctly. Hence, the probability of Eve
executing a successful impersonation attack is $\left(\frac{1}{4}\right)^{n}$.
As $n$ grows large, the likelihood of a successful impersonation
attack diminishes nearly to zero. Consequently, the probability $P\left(n\right)$
of detecting Eve's presence is $1-\left(\frac{1}{4}\right)^{n}$.
For sufficiently large $n$ values, $P\left(n\right)$ approximates
$1$, facilitating the identification of an impersonation attack without
failure. The relationship between $P\left(n\right)$ and $n$ is depicted
in Fig. \ref{fig:Figure_2}, which illustrates the necessity of at
least 6 pre-shared keys to detect Eve's presence effectively.

\subsection{Security assessment of the protocol against intercept and resend
attack}

\begin{figure}
\begin{centering}
\includegraphics[scale=0.5]{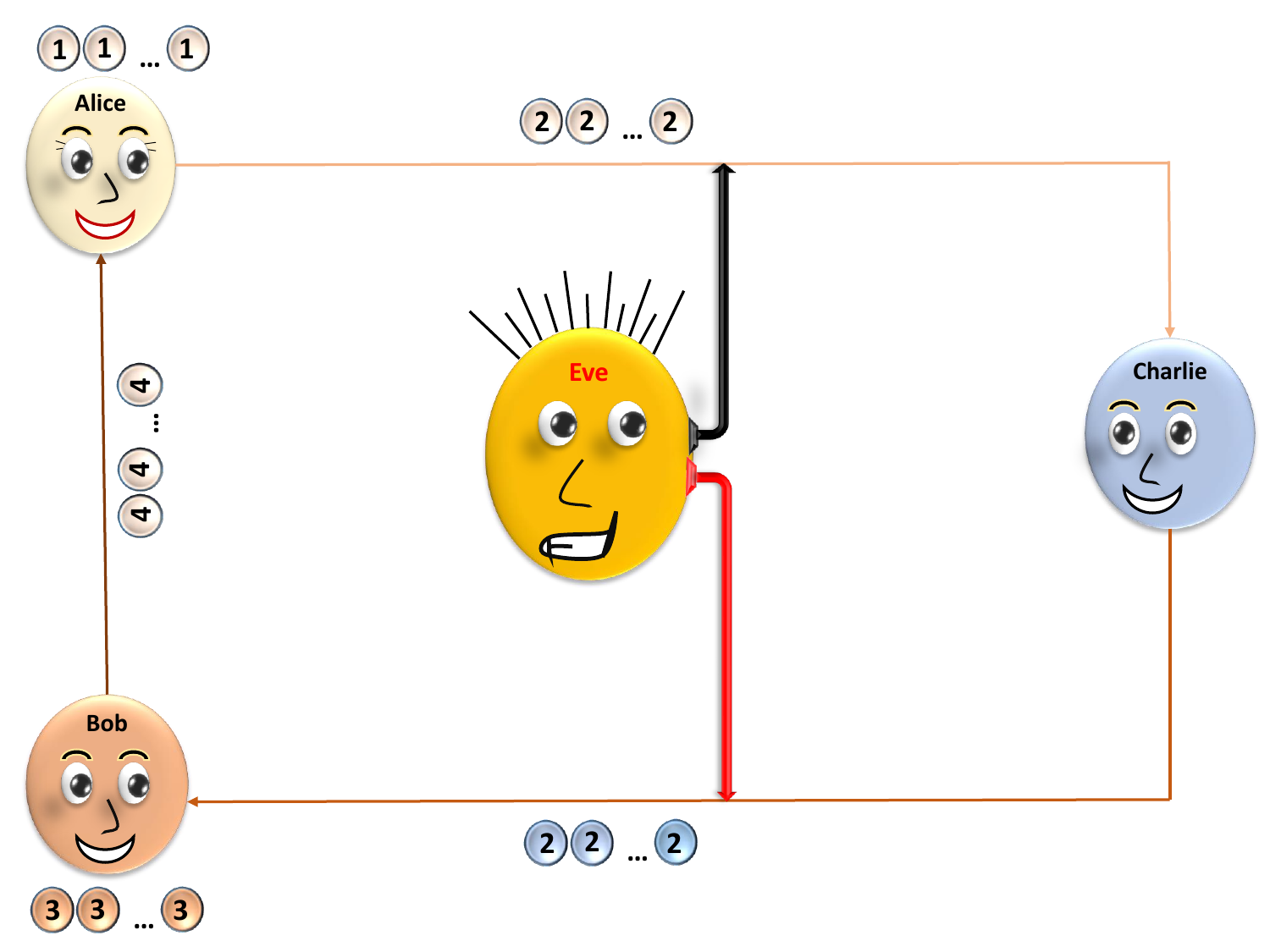}
\par\end{centering}
\caption{\label{fig:Figure_3}Eve\textquoteright s IR attack strategy. Here,
the numbers 1, 2, 3 and 4 represent the qubits in sequences $S_{A1}$,
$S_{A2}$, $S_{B3}$ and $S_{B4}$, respectively.}
\end{figure}

In this quantum protocol, Alice and Bob generate Bell states based
on their pre-shared secret key. Instead of transmitting their entire quantum states, Alice and Bob selectively send Particle
2 and Particle 4 through the quantum channel. This approach limits
Eve's potential attacks to these specific particles at any given moment.
For the sake of simplicity, we assume that Eve will attempt to intercept
Particle 2 and Particle 4 as they are transmitted from Alice to Charlie
and from Bob to Alice, respectively. Eve's ability to extract information
from the quantum channel is constrained by the Holevo bound or Holevo
quantity \cite{H_73} 

\begin{equation}
\begin{array}{lcl}
\chi\left(\rho\right) & = & S\left(\rho\right)-\underset{i}{\sum}{\rm p}_{i}S\left(\rho_{i}\right),\end{array}\label{eq:Holevo Information}
\end{equation}
which determines the maximum amount of information she can obtain.
In the given context, $S\left(\rho\right)=-{\rm Tr}\left(\rho\,{\rm log_{2}}\rho\right)$
represents the von Neumann entropy, where $\rho_{i}$ denotes a component
within the mixed state $\rho$ with probability ${\rm p}_{i}$. The
expression for $\rho$ is formulated as $\rho=\frac{1}{4}\left[\left|\Psi^{00}\right\rangle +\left|\Psi^{01}\right\rangle +\left|\Psi^{10}\right\rangle +\left|\Psi^{11}\right\rangle \right]_{1234}=\underset{i}{\sum\,}{\rm p}_{i}\rho_{i}$,
where $\rho_{{\rm i}}$ corresponds to the density matrix of the states
mentioned in Eqs. (\ref{eq:Final composite system=00007B11=00007D}),
(\ref{eq:Final composite system=00007B00=00007D}), (\ref{eq:Final composite system=00007B01=00007D})
and (\ref{eq:Final composite system=00007B10=00007D}). Earlier, it
was established that Eve's objective is to target particles 2 and
4. We aim to illustrate a scenario where Eve can successfully attack
both particles 2 and 4 simultaneously. In this context, we analyze
that Eve's ability to gain any meaningful information is limited by
the Holevo quantity. To align with this requirement, we adjust Eq.
(\ref{eq:Holevo Information}) as follows:

\begin{equation}
\begin{array}{lcl}
\chi\left(\rho^{24}\right) & = & S\left(\rho^{24}\right)-\underset{i}{\sum}{\rm p}_{i}S\left(\rho_{i}^{24}\right),\end{array}\label{eq:Holevo Information-1}
\end{equation}
where $\rho^{24}$ and $\rho_{i}^{24}$ represent the reduced matrices
of $\rho$ and $\rho_{i}$, respectively, obtained after performing
a partial trace over particles 1 and 3. Through straightforward calculations,
we determine that $\rho^{24}={\rm Tr}_{13}\left(\rho\right)=\frac{1}{4}\left(|11\rangle\langle11|+|10\rangle\langle10|+|01\rangle\langle01|+|00\rangle\langle00|\right)=\frac{1}{4}\mathds{1}_{4}$,
where $\mathds{1}_{4}$ denotes the $4\times4$ identity matrix. Similarly,
the von Neumann entropy of each component of the mixed state $\rho$
after the partial trace, denoted as $\rho_{i}^{24}={\rm Tr}_{13}\left(\rho_{i}\right)=\frac{1}{4}\mathds{1}_{4}$.
By substituting $\rho^{24}$ and $\rho_{i}^{24}$ into Eq. (\ref{eq:Holevo Information-1}),
we deduce that $\chi\left(\rho^{24}\right)=0$. Consequently, we infer
that Eve cannot obtain any key information through direct intercept
attacks on the transmission particles. The intercept-resend (IR) attack
strategy employed by Eve is depicted in Fig. \ref{fig:Figure_3}. The maximally mixed
state is considered with equal probabilities of Bell states in Eq.
(\ref{eq:Holevo Information}) to minimize or eliminate Eve's
information gain from the channel in IR attack. If these probabilities
are not equal, the security against the IR attack will depend on the
specific values of the Bell state probabilities in the mixed state.
As long as the mutual information between Alice and Bob remains greater
than the Holevo information $\left(I(A|B)>\chi\left(\rho\right)\right)$,
the scheme will remain secure against the IR attack \cite{CRE_04}.

\subsection{Security assessment of the protocol against impersonated fraudulent
attack}

To demonstrate our protocol against impersonated fraudulent attack,
we consider two pre-shared secret keys as 11 and 00, respectively.
Initially, Alice and Bob prepare the Bell state $|\phi^{+}\rangle_{12}$
and $|\psi^{-}\rangle_{34}$, respectively. Meanwhile, Eve prepares
two new single-qubit states $|\chi\rangle_{5}$ and $|\chi\rangle_{6}$
as fake states, which are described as $a|0\rangle+b|1\rangle$ and
$c|0\rangle+d|1\rangle$, respectively, where $\left|a\right|^{2}+\left|b\right|^{2}=1$
and $\left|c\right|^{2}+\left|d\right|^{2}=1$. Eve then performs
a CNOT operation with particle 2 (4) acting as the control qubit and
the target qubit being particle 5 (6). She retains particles 2 and
4, and sends particles 5 and 6 to Bob and Alice, respectively. The
composite state following Eve's attack, as induced by the CNOT operation,
is expressed as follows:

\begin{equation}
\begin{array}{lcl}
 &  & {\rm CNOT_{2(4)\rightarrow5(6)}|\phi^{+}\rangle_{12}\otimes|\psi^{-}\rangle_{34}\otimes|\chi\rangle_{5}\otimes|\chi\rangle_{6}}\\
 & = & \frac{1}{2}\left[|0001\rangle\left(a|0\rangle+b|1\rangle\right)\left(c|1\rangle+d|0\rangle\right)-|0010\rangle\left(a|0\rangle+b|1\rangle\right)\left(c|0\rangle+d|1\rangle\right)\right.\\
 & + & \left.|1101\rangle\left(a|1\rangle+b|0\rangle\right)\left(c|1\rangle+d|0\rangle\right)-|1110\rangle\left(a|1\rangle+b|0\rangle\right)\left(c|0\rangle+d|1\rangle\right)\right]_{123456}\\
 & = & \frac{1}{2}\left[ac|000101\rangle+ad|000100\rangle+bc|000111\rangle+bd|000110\rangle\right.\\
 & - & ac|001000\rangle-ad|001001\rangle-bc|001010\rangle-bd|001011\rangle\\
 & + & ac|110111\rangle-ad|110110\rangle-bc|110101\rangle-bd|110100\rangle\\
 & - & \left.ac|111010\rangle-ad|111011\rangle-bc|111000\rangle-bd|111001\rangle\right]_{123456}
\end{array}.\label{eq:CNOT_SingleQubit}
\end{equation}
As per the protocol, Alice and Bob execute $\sigma_{z}$ operation
on their respective particles 1 and 3. Subsequently, they conduct
Bell measurements on particles 1 and 6 for Alice, and on particles
5 and 3 for Bob. Eve is assigned particles 2 and 4. The resultant
composite system of Alice, Bob and Eve, subsequent to the Pauli operation
and Bell measurement described in Eq. (\ref{eq:CNOT_SingleQubit}),
can be represented as:

\begin{equation}
\begin{array}{lcl}
|\Psi\rangle & = & \frac{1}{2\sqrt{2}}\left[ac\left(|\psi^{+}\rangle|\phi^{+}\rangle|\psi^{-}\rangle+|\psi^{+}\rangle|\phi^{-}\rangle|\psi^{+}\rangle+|\psi^{-}\rangle|\phi^{+}\rangle|\psi^{+}\rangle+|\psi^{-}\rangle|\phi^{-}\rangle|\psi^{+}\rangle\right.\right.\\
 & + & \left.|\phi^{+}\rangle|\psi^{+}\rangle|\phi^{-}\rangle+|\phi^{+}\rangle|\psi^{-}\rangle|\phi^{+}\rangle+|\phi^{-}\rangle|\psi^{+}\rangle|\phi^{+}\rangle+|\phi^{-}\rangle|\psi^{-}\rangle|\phi^{-}\rangle\right)\\
 & + & ad\left(|\phi^{+}\rangle|\phi^{+}\rangle|\psi^{-}\rangle+|\phi^{+}\rangle|\phi^{-}\rangle|\psi^{+}\rangle+|\phi^{-}\rangle|\phi^{+}\rangle|\psi^{+}\rangle+|\phi^{-}\rangle|\phi^{-}\rangle|\psi^{-}\rangle\right.\\
 & + & \left.|\psi^{+}\rangle|\psi^{+}\rangle|\phi^{-}\rangle+|\psi^{+}\rangle|\psi^{-}\rangle|\phi^{+}\rangle+|\psi^{-}\rangle|\psi^{+}\rangle|\phi^{+}\rangle+|\psi^{-}\rangle|\psi^{-}\rangle|\phi^{-}\rangle\right)\\
 & + & bc\left(|\psi^{+}\rangle|\psi^{+}\rangle|\psi^{-}\rangle-|\psi^{+}\rangle|\psi^{-}\rangle|\psi^{+}\rangle+|\psi^{-}\rangle|\psi^{+}\rangle|\psi^{+}\rangle-|\psi^{-}\rangle|\psi^{-}\rangle|\psi^{-}\rangle\right.\\
 & + & \left.|\phi^{+}\rangle|\phi^{+}\rangle|\phi^{-}\rangle-|\phi^{+}\rangle|\phi^{-}\rangle|\phi^{+}\rangle+|\phi^{-}\rangle|\phi^{+}\rangle|\phi^{+}\rangle-|\phi^{-}\rangle|\phi^{-}\rangle|\phi^{-}\rangle\right)\\
 &  & bd\left(|\phi^{+}\rangle|\psi^{+}\rangle|\psi^{-}\rangle-|\phi^{+}\rangle|\psi^{-}\rangle|\psi^{+}\rangle+|\phi^{-}\rangle|\psi^{+}\rangle|\psi^{+}\rangle-|\phi^{-}\rangle|\psi^{-}\rangle|\psi^{-}\rangle\right.\\
 & + & \left.\left.|\psi^{+}\rangle|\phi^{+}\rangle|\phi^{-}\rangle-|\psi^{+}\rangle|\phi^{-}\rangle|\phi^{+}\rangle+|\psi^{-}\rangle|\phi^{+}\rangle|\phi^{+}\rangle-|\psi^{-}\rangle|\phi^{-}\rangle|\phi^{-}\rangle\right)\right]_{165324}
\end{array}.\label{eq:Final_Composite_State_Single_Qubit}
\end{equation}
When Alice and Bob obtain Bell pairs in the states $\left|\psi^{+}\right\rangle _{16}\left|\phi^{-}\right\rangle _{53},\left|\psi^{-}\right\rangle _{16}\left|\phi^{+}\right\rangle _{53},\left|\phi^{+}\right\rangle _{16}\left|\psi^{-}\right\rangle _{53}\,{\rm and }\left|\phi^{-}\right\rangle _{16}\left|\psi^{+}\right\rangle _{53}$
after measurement, detection probability of eavesdropper's presence
becomes zero. By analyzing the density operator of the final composite
state $|\Psi\rangle\langle\Psi|$, as from Eq. (\ref{eq:Final_Composite_State_Single_Qubit}),
we can determine the non-detection probability of Eve's presence,
denoted as ${\rm P_{nd}}$, to be $\frac{1}{2}\left(\left|ac\right|^{2}+\left|bd\right|^{2}\right)$.
Therefore, the detection probability of Eve's presence, denoted as
${\rm P_{d}}$, is ${\rm P_{d}=1-}\frac{1}{2}\left(\left|ac\right|^{2}+\left|bd\right|^{2}\right)$.
Eve can minimize her detection probability by setting $\left|a\right|=\left|b\right|=\left|c\right|=\left|d\right|=\frac{1}{\sqrt{2}}$.
Thus, the single qubit states of Eve are $|+\rangle_{5}=\frac{1}{\sqrt{2}}\left(|0\rangle+|1\rangle\right)_{5}$
and $|+\rangle_{6}=\frac{1}{\sqrt{2}}\left(|0\rangle+|1\rangle\right)_{6}$.
Under this condition, ${\rm P_{d}}=\frac{3}{4}$, representing the
minimum detection probability of Eve when she selects the single qubit
state as a fake state.

Now, we investigate the same scenario with the only difference being
Eve's entangled state as her fake state to impersonate herself as
a legitimate party. In this scenario, we consider the same pair of
pre-shared keys, $11,00$, and Eve's fake state is $|\chi^{\prime}\rangle_{56}=\left(a^{\prime}|00\rangle+b^{\prime}|01\rangle+c^{\prime}|10\rangle+d^{\prime}|11\rangle\right)_{56}$,
where $\left|a^{\prime}\right|^{2}+\left|b^{\prime}\right|^{2}+\left|c^{\prime}\right|^{2}+\left|d^{\prime}\right|^{2}=1$.
The remaining process is the same as the previous one. Eventually,
the final composite system of Alice, Bob and Eve after their Bell
measurement is 

\begin{equation}
\begin{array}{lcl}
|\Psi^{\prime}\rangle & = & \frac{1}{2\sqrt{2}}\left[a^{\prime}\left(|\psi^{+}\rangle|\phi^{+}\rangle|\psi^{-}\rangle+|\psi^{+}\rangle|\phi^{-}\rangle|\psi^{+}\rangle+|\psi^{-}\rangle|\phi^{+}\rangle|\psi^{+}\rangle+|\psi^{-}\rangle|\phi^{-}\rangle|\psi^{+}\rangle\right.\right.\\
 & + & \left.|\phi^{+}\rangle|\psi^{+}\rangle|\phi^{-}\rangle+|\phi^{+}\rangle|\psi^{-}\rangle|\phi^{+}\rangle+|\phi^{-}\rangle|\psi^{+}\rangle|\phi^{+}\rangle+|\phi^{-}\rangle|\psi^{-}\rangle|\phi^{-}\rangle\right)\\
 & + & b^{\prime}\left(|\phi^{+}\rangle|\phi^{+}\rangle|\psi^{-}\rangle+|\phi^{+}\rangle|\phi^{-}\rangle|\psi^{+}\rangle+|\phi^{-}\rangle|\phi^{+}\rangle|\psi^{+}\rangle+|\phi^{-}\rangle|\phi^{-}\rangle|\psi^{-}\rangle\right.\\
 & + & \left.|\psi^{+}\rangle|\psi^{+}\rangle|\phi^{-}\rangle+|\psi^{+}\rangle|\psi^{-}\rangle|\phi^{+}\rangle+|\psi^{-}\rangle|\psi^{+}\rangle|\phi^{+}\rangle+|\psi^{-}\rangle|\psi^{-}\rangle|\phi^{-}\rangle\right)\\
 & + & c^{\prime}\left(|\psi^{+}\rangle|\psi^{+}\rangle|\psi^{-}\rangle-|\psi^{+}\rangle|\psi^{-}\rangle|\psi^{+}\rangle+|\psi^{-}\rangle|\psi^{+}\rangle|\psi^{+}\rangle-|\psi^{-}\rangle|\psi^{-}\rangle|\psi^{-}\rangle\right.\\
 & + & \left.|\phi^{+}\rangle|\phi^{+}\rangle|\phi^{-}\rangle-|\phi^{+}\rangle|\phi^{-}\rangle|\phi^{+}\rangle+|\phi^{-}\rangle|\phi^{+}\rangle|\phi^{+}\rangle-|\phi^{-}\rangle|\phi^{-}\rangle|\phi^{-}\rangle\right)\\
 &  & d^{\prime}\left(|\phi^{+}\rangle|\psi^{+}\rangle|\psi^{-}\rangle-|\phi^{+}\rangle|\psi^{-}\rangle|\psi^{+}\rangle+|\phi^{-}\rangle|\psi^{+}\rangle|\psi^{+}\rangle-|\phi^{-}\rangle|\psi^{-}\rangle|\psi^{-}\rangle\right.\\
 & + & \left.\left.|\psi^{+}\rangle|\phi^{+}\rangle|\phi^{-}\rangle-|\psi^{+}\rangle|\phi^{-}\rangle|\phi^{+}\rangle+|\psi^{-}\rangle|\phi^{+}\rangle|\phi^{+}\rangle-|\psi^{-}\rangle|\phi^{-}\rangle|\phi^{-}\rangle\right)\right]_{165324}
\end{array}.\label{eq:Final_Composite_State_Entangled_State}
\end{equation}
In the scenario described, the detection probability of Eve's presence
is null when Alice and Bob obtain measurement results associated with
specific Bell pairs: $\left|\psi^{+}\right\rangle _{16}\left|\phi^{-}\right\rangle _{53},\left|\psi^{-}\right\rangle _{16}\left|\phi^{+}\right\rangle _{53},\left|\phi^{+}\right\rangle _{16}\left|\psi^{-}\right\rangle _{53}\,{\rm and }\left|\phi^{-}\right\rangle _{16}\left|\psi^{+}\right\rangle _{53}$.
The non-detection probability of Eve's presence can be computed from
the density matrix, $|\Psi^{\prime}\rangle\langle\Psi^{\prime}|$
from Eq. (\ref{eq:Final_Composite_State_Entangled_State}), denoted
as ${\rm P_{nd}}=\frac{1}{2}\left(\left|a^{\prime}\right|^{2}+\left|d^{\prime}\right|^{2}\right)$.
Correspondingly, the detection probability of Eve's presence, denoted
as ${\rm P_{d}}=1-\frac{1}{2}\left(\left|a^{\prime}\right|^{2}+\left|d^{\prime}\right|^{2}\right)$.
Eve strategically sets $\left|a^{\prime}\right|=\left|d^{\prime}\right|=\frac{1}{\sqrt{2}}$
and $\left|b^{\prime}\right|=\left|c^{\prime}\right|=0$ to minimize
her detection probability. The entangled state of Eve is represented
by $|\phi^{+}\rangle_{56}=\frac{1}{\sqrt{2}}\left(|00\rangle+|11\rangle\right)_{56}$,
resulting in ${\rm P_{d}}=\frac{1}{2}$. It follows that using an
entangled state rather than a single qubit state reduces the detection
probability of Eve's presence. Consequently, the minimum detection
probabilities for single qubit and entangled states are $\frac{3}{4}$
and $\frac{1}{2}$ respectively. Based on the discussion, it can be
deduced that the protocol is secure against impersonated fraudulent
attack by Eve, particularly under optimal conditions for Eve.

\section{Collective noise analysis of proposed protocol\protect\label{sec:IV}}

The QIA protocol
mentioned earlier was discussed assuming an ideal quantum channel.
In practice, when particles are transmitted through a quantum channel,
they are exposed to noise, which can alter their state(s). Additionally,
an attacker might exploit the presence of noise to mask his attack(s),
making it difficult to distinguish between errors caused by noise
and those introduced by an attacker. Addressing the impact of collective
noise on quantum communication is a well-known challenge \cite{CGL+18,HM16,HWL++14}.
Walton et al. \cite{WAS+03} introduced the concept of a decoherence-free
subspace (DFS), which can mitigate the effects of collective noise
since the states of the particles remain unaffected within these channels.
Further research has also been conducted on quantum communication
under collective noise \cite{HM17,GCQ18}. Collective noise typically
includes both collective-dephasing and collective-rotation noise \cite{LDZ08}.
In what follows, we will analyze the impact of these types of noise
on the scheme proposed here.

\subsection{Collective-dephasing noise}

A collective-dephasing noise can be expressed as
\cite{LDZ08,GCQ18}

\[
\begin{array}{lcl}
U_{dp}|0\rangle=|0\rangle & , & U_{dp}|1\rangle=e^{i\phi}|1\rangle\end{array}.
\]
Collective-dephasing noise is characterized by a parameter $\phi$,
which varies with time. Typically, a logical qubit encoded in the
product states of two physical qubits as 
\[
\begin{array}{lclc}
|0\rangle_{L}=|01\rangle & , & |1\rangle_{L}=|10\rangle & ,\end{array}
\]
is resistant to this type of noise because both logical qubits accumulate
the same phase factor $e^{i\phi}$.

\begin{figure}[h]
\begin{centering}
\includegraphics[scale=0.5]{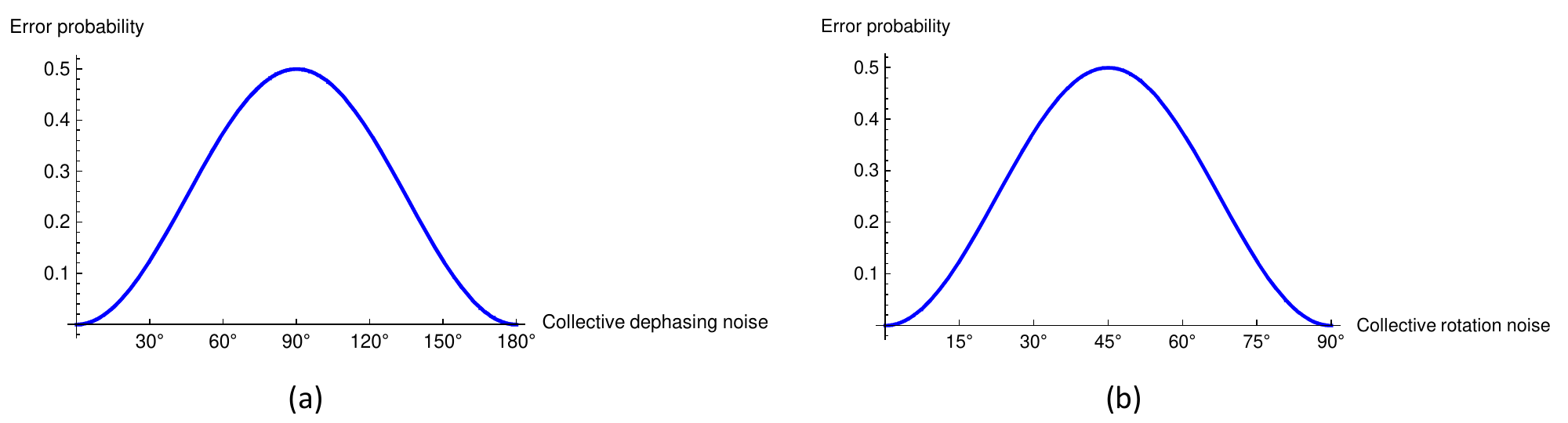}
\par\end{centering}
\caption{\protect\label{fig:Collective-error-probability}The collective error
probability with respect to the noise parameter can be categorized
as: (a) Collective dephasing error probability with noise parameter
$\phi$, and (b) Collective rotation error probability with noise
parameter $\theta$.}

\end{figure}

In this context, the subscript $L$ denotes the logical
qubit, while 0 and 1 correspond to horizontal and vertical polarization
states, respectively; these being the eigenstates of the Pauli operator
$\sigma_{z}$ ($Z$ basis). The states $|+_{L}\rangle$ and
$|-_{L}\rangle$ are defined as follows:

\[
\begin{array}{ccccc}
|+_{L}\rangle & = & \frac{1}{\sqrt{2}}\left(|0_{L}\rangle+|1_{L}\rangle\right) & = & \frac{1}{\sqrt{2}}\left(|01\rangle+|10\rangle\right)\\
|-_{L}\rangle & = & \frac{1}{\sqrt{2}}\left(|0_{L}\rangle-|1_{L}\rangle\right) & = & \frac{1}{\sqrt{2}}\left(|01\rangle+|10\rangle\right)
\end{array}.
\]
From the equations presented in Section \ref{subsec:Principal-concept},
specifically from (\ref{eq:Final composite system=00007B11=00007D})
and (\ref{eq:Final composite system=00007B10=00007D}),
it is evident that particles 2 and 4 in the composite states are responsible
for transmitting quantum information through the quantum channel.
Taking Eq. (\ref{eq:Final composite system=00007B11=00007D})
as an example:

\[
\begin{array}{lcl}
\left|\Psi^{11}\right\rangle  & = & \sigma_{z1}\otimes\sigma_{z3}\left(\left|\phi^{+}\right\rangle _{12}\left|\psi^{-}\right\rangle _{34}\right)\\
 & = & \frac{1}{2}\sigma_{z1}\otimes\sigma_{z3}\left(|0001\rangle-|0010\rangle+|1101\rangle-|1110\rangle\right)_{1234}\\
 & = & \frac{1}{2}\left(|0001\rangle+|0010\rangle-|1101\rangle-|1110\rangle\right)_{1234}
\end{array},
\]
after considering collective-dephasing noise on particles $2$ and
$4$, we obtain

\[
\begin{array}{lcl}
\left|\Psi^{11}\right\rangle _{dp} & = & \frac{1}{2}\left(e^{i\phi}|0001\rangle+|0010\rangle-e^{2i\phi}|1101\rangle-e^{i\phi}|1110\rangle\right)_{1234}\\
 & = & \frac{1}{2}\left(e^{i\phi}|01\rangle|00\rangle+|00\rangle|01\rangle-e^{2i\phi}|11\rangle|10\rangle-e^{i\phi}|10\rangle|11\rangle\right)_{1423}\\
 & = & \frac{1}{4}\left[2e^{i\phi}\left(|\psi^{+}\rangle|\phi^{-}\rangle+|\psi^{-}\rangle|\phi^{+}\rangle\right)+\left(1-e^{2i\phi}\right)\left(|\phi^{+}\rangle|\psi^{+}\rangle+|\phi^{-}\rangle|\psi^{-}\rangle\right)_{1423}\right..\\
 & + & \left.\left(1+e^{2i\phi}\right)\left(|\phi^{+}\rangle|\psi^{-}\rangle+|\phi^{-}\rangle|\psi^{+}\rangle\right)\right]_{1423}
\end{array}
\]
If there are no errors in the channel, the output Bell pairs shared
between Alice and Bob would be from the set $\left\{ |\psi^{-}\rangle|\phi^{+}\rangle,|\psi^{+}\rangle|\phi^{-}\rangle,|\phi^{-}\rangle|\psi^{+}\rangle,|\phi^{+}\rangle|\psi^{-}\rangle\right\} $.
However, due to collective dephasing noise, the output pairs may become
$\left\{ |\phi^{+}\rangle|\psi^{+}\rangle,|\phi^{-}\rangle|\psi^{-}\rangle\right\} $.
Such situations will introduce errors. The probability of an error
caused by this noise is computed as $\frac{1}{4}\left(1-\cosh(2i\phi)\right)$,
where $\phi$ represents the noise parameter. Figure \ref{fig:Collective-error-probability}
illustrates this collective error probability as a function of the
noise parameter. According to Fig. \ref{fig:Collective-error-probability}.(a),
the error probability follows a curve, reaching a maximum of 0.5 when
$\phi=90^{\circ}$, and approaching zero at $\phi=0^{\circ}\,{\rm and}\,180^{\circ}$.
To reduce or circumvent the error induced by this type of noise, we
should select a channel that keeps the error with in tolerable error
limit. Specifically, if tolerable error limit is computed as $p$
then we have to find out solutions of $p=\frac{1}{4}\left(1-\cosh(2i\phi)\right)$
for $0^{\circ}<\phi<180^{\circ}.$ If we obtain two solutions of it
as $\phi_{1}$ and $\phi_{2}$ such that $\phi_{1}<\phi_{2},$ then
the protocol would remain secure in a channel where $\phi<\phi_{1}$
or $\phi>\phi_{2}.$ To know whether that's the case in a practical
scenario, channel characterization must be performed before implementing
the protocol. 

\subsection{Collective-rotation noise}

The collective-rotation noise can be represented
as:

\[
\begin{array}{lcl}
U_{r}|0\rangle=\cos\theta|0\rangle+\sin\theta|1\rangle & , & U_{r}|1\rangle=-\sin\theta|0\rangle+\cos\theta|1\rangle\end{array},
\]
The collective rotation noise is represented by the parameter $\theta$,
which varies over time in the quantum channel. For example, applying
collective rotation noise to particles 2 and 4 from Eq. (\ref{eq:Final composite system=00007B11=00007D})
results in the following expression after a tedious calculations,

\[
\begin{array}{lcl}
U_{r2}\otimes U_{r4}\left|\Psi^{11}\right\rangle  & = & \frac{1}{2}\left[-\sin\theta\cos\theta|0000\rangle-\sin^{2}\theta|0010\rangle+\cos^{2}\theta|0100\rangle+\sin\theta\cos\theta|0110\rangle\right.\\
 & + & \sin\theta\cos\theta|1001\rangle-\cos^{2}\theta|1011\rangle+\sin^{2}\theta|1101\rangle-\sin\theta\cos\theta|1111\rangle\\
 & - & \left(\cos^{2}\theta|0001\rangle+\sin\theta\cos\theta|0011\rangle+\sin\theta\cos\theta|0101\rangle+\sin^{2}\theta|0111\rangle\right.\\
 & - & \left.\left.\sin^{2}\theta|1000\rangle+\sin\theta\cos\theta|1010\rangle+\sin\theta\cos\theta|1100\rangle-\cos^{2}\theta|1110\rangle\right)\right]_{1423}.\\
 & = & \frac{1}{2}\left[2\sin\theta\cos\theta\left(-|\phi^{+}\rangle|\phi^{+}\rangle-|\psi^{-}\rangle|\psi^{-}\rangle\right)\right.\\
 & + & \sin^{2}\theta\left(|\phi^{+}\rangle|\psi^{-}\rangle-|\phi^{-}\rangle|\psi^{+}\rangle+|\psi^{+}\rangle|\phi^{-}\rangle-|\psi^{-}\rangle|\phi^{+}\rangle\right)\\
 & + & \left.\cos^{2}\theta\left(|\psi^{+}\rangle|\phi^{-}\rangle+|\psi^{-}\rangle|\phi^{+}\rangle-|\phi^{+}\rangle|\psi^{-}\rangle-|\phi^{-}\rangle|\psi^{+}\rangle\right)\right]_{1423}
\end{array}
\]
In the absence of noise or errors, the Bell pairs shared by Alice
and Bob are $\left\{ |\psi^{-}\rangle|\phi^{+}\rangle,|\psi^{+}\rangle|\phi^{-}\rangle,|\phi^{-}\rangle|\psi^{+}\rangle,|\phi^{+}\rangle|\psi^{-}\rangle\right\} $.
From this above result, the probability of error due to collective
rotation noise can be derived as $2\sin^{2}\theta\cos^{2}\theta$,
which depends on the noise parameter $\theta$. Similarly, applying
collective rotation noise to particles 2 and 4 from Eq. (\ref{eq:Final composite system=00007B00=00007D})
yields: 
\[
\begin{array}{lcl}
U_{r2}\otimes U_{r4}\left|\Psi^{00}\right\rangle  & = & \frac{1}{2}\left[\cos^{2}\theta|0000\rangle+\sin\theta\cos\theta|0010\rangle+\sin\theta\cos\theta|0100\rangle+\sin^{2}\theta|0110\rangle\right.\\
 & + & \sin^{2}\theta|1001\rangle-\sin\theta\cos\theta|1011\rangle-\sin\theta\cos\theta|1101\rangle+\cos^{2}\theta|1111\rangle\\
 & - & \left(-\sin\theta\cos\theta|0001\rangle-\sin^{2}\theta|0011\rangle+\cos^{2}\theta|0101\rangle+\sin\theta\cos\theta|0111\rangle\right.\\
 & - & \left.\left.-\sin\theta\cos\theta|1000\rangle+\cos^{2}\theta|1010\rangle-\sin^{2}\theta|1100\rangle+-\sin\theta\cos\theta|1110\rangle\right)\right]_{1423}.\\
 & = & \frac{1}{2}\left[2\sin\theta\cos\theta\left(|\phi^{-}\rangle|\psi^{+}\rangle+|\psi^{+}\rangle|\phi^{-}\rangle\right)\right.\\
 & + & \sin^{2}\theta\left(|\psi^{+}\rangle|\psi^{+}\rangle-|\psi^{-}\rangle|\psi^{-}\rangle+|\phi^{+}\rangle|\phi^{+}\rangle-|\phi^{-}\rangle|\phi^{-}\rangle\right)\\
 & + & \left.\cos^{2}\theta\left(|\phi^{+}\rangle|\phi^{+}+|\phi^{-}\rangle|\phi^{-}\rangle+|\psi^{+}\rangle|\psi^{+}\rangle+|\psi^{-}\rangle|\psi^{-}\rangle\right)\right]_{1423}
\end{array}
\]
For the state $|\Psi^{00}\rangle$, no error occurs if the final Bell
pairs shared by Alice and Bob are $\left\{ |\phi^{+}\rangle|\phi^{+}\rangle,|\phi^{-}\rangle|\phi^{-}\rangle,|\psi^{+}\rangle|\psi^{+}\rangle,|\psi^{-}\rangle|\psi^{-}\rangle\right\} $.
By analyzing the above result, it is evident that the error probability
due to collective rotation noise follows the same form, $2\sin^{2}\theta\cos^{2}\theta$,
as in the case of $|\Psi^{11}\rangle$. The error probability is plotted
as a function of $\theta$ in Fig. \ref{fig:Collective-error-probability}.(b).
From Fig. \ref{fig:Collective-error-probability}.(b), the rotation
error probability reaches its maximum value (0.5) when $\theta=45^{\circ}$,
and it minimizes to zero at $\theta=0^{\circ}\,{\rm and}\,90^{\circ}$.
To minimize or eliminate the error caused by this type of noise, it
is essential to select a channel that maintains the error within a
tolerable limit. Specifically, if the tolerable error limit is denoted
as $p^{\prime}$, we must solve the equation $p^{\prime}=2\sin^{2}\theta\cos^{2}\theta$
for $0^{\circ}<\theta<90^{\circ}$. If the solutions are $\theta_{1}$
and $\theta_{2}$ with $\theta_{1}<\theta_{2},$ the protocol will
remain secure in a channel where $\theta<\theta_{1}$ or $\theta>\theta_{2}.$
To verify if this condition holds in a practical setting, channel
characterization should be conducted prior to protocol implementation.
Recent advancements in quantum communication protocols demonstrate
inherent resistance to collective noise \cite{TFI+16,WCS22,LWZ+16}.
Similar techniques can be applied to our scheme to enhance resistance
to collective noise. 

\section{Comparison with the existing protocols\label{sec:V}}

In this section, we aim to conduct a concise comparative analysis
involving the proposed protocol and a selection of previously proposed
QIA schemes. The comparison focuses on several key aspects: the utilization
of quantum resources, the nature of the third-party involvement (whether
they are assumed honest/semi-honest or considered untrusted), the
minimum number of pre-shared keys required for authentication, and
whether the scheme enables bidirectional mutual authentication or
only allows unidirectional authentication where it permits only one
legitimate party to check the authenticity of the other. Given the
abundance of QIA schemes, we have carefully chosen representative
ones for comparison, with particular attention to schemes based on
protocols for secure direct quantum communication, akin to the present
protocol, which draws inspiration from CDSQC.

We commence the comparison by examining Zhang et al.'s QIA scheme
\cite{ZZZX_2006}, which relied on the ping-pong protocol for QSDC.
In Zhang et al.'s protocol, Alice serves as the reliable certification
authority, and Bob represents the common user whose identity requires
verification by Alice. Similarly, Yuan et al.'s protocol \cite{YLP+14},
based on the LM05 protocol for single photon-based QSDC, follows a
similar structure with Alice acting as the certification authority.
Both of these QSDC-based protocols for QIA \cite{ZZZX_2006,YLP+14}
are unidirectional, unlike the proposed bidirectional scheme. This
highlights an advantage of the proposed protocol over in \cite{ZZZX_2006,YLP+14}.
Notably, this advantage extends over a range of unidirectional QIA
protocols, such as Hong et al.'s protocol \cite{HCJ+17}. However,
it is important to note that this advantage is not unique, as similar
bidirectional capabilities are also found in Kang et al.'s protocols
\cite{KHHYHM_2018,KHH+20} and in Zhang et al.'s 2020 work \cite{ZCSL_2020}.
Remarkably, all these schemes \cite{ZCSL_2020,KHHYHM_2018,KHH+20}
necessitate a minimum of 6 pre-shared keys for authentication, whereas
our protocol requires only 4, which can be seen as advantageous due
to the fact that the quantum resources are costly. Moreover, Kang
et al.'s works \cite{KHHYHM_2018,KHH+20} utilize GHZ-like states,
which are tripartite entangled states, posing challenges in preparation
and maintenance compared to the Bell states utilized in our protocol.
Additionally, Zhang et al.'s 2020 work assumes the third party to
be semi-honest, whereas our proposed scheme considers the third party
as untrusted, enhancing security. Jiang et al. proposed a semi-quantum
QIA scheme resembling the ping-pong protocol \cite{JZH21}, where
Bob possesses only classical capabilities with limited quantum access,
while Alice has more extensive quantum resources to prepare and measure
Bell states, among other quantum operations. However, this setup introduces
increased noise probability due to two-way communication and full
access for Eve over the Bell state in the quantum channel. In contrast,
the proposed scheme utilizes a one-way quantum channel and restricts
Eve from accessing the complete Bell state. Another endeavor in QIA
with QKA utilizing Bell and GHZ states with semi-honest third-party
involvement was presented by Wu et al. \cite{WCG+21}. However, their
scheme demands significant quantum resources and poses maintenance
challenges compared to our proposed scheme. Furthermore, their use
of hash functions for authentication may not align with quantum security
principles. Similarly, Li et al. proposed a simultaneous QIA scheme
using GHZ states \cite{LZZ+22}, but maintaining such quantum states
proves difficult compared to Bell states. Thus, our proposed scheme
demonstrates the desired QIA features while optimizing resource utilization,
particularly when compared to the existing entangled state-based QIA
schemes. The comparative analysis presented herein aims to underscore
the relevance and advantages of the proposed protocol. A summarized
comparison is provided in Table \ref{tab:Comparison} for clarity.

\begin{table}
\begin{centering}
\begin{tabular*}{16.5cm}{@{\extracolsep{\fill}}>{\centering}p{2cm}>{\centering}p{2.5cm}>{\centering}p{2cm}c>{\centering}p{2cm}>{\centering}p{4cm}}
\hline 
 &  &  &  &  & \tabularnewline
Protocol & Quantum resources & Minimum secret key required & Way of authentication & Nature of the third party & Secure against attacks\tabularnewline
 &  &  &  &  & \tabularnewline
\hline 
 &  &  &  &  & \tabularnewline
Zhang et al. \cite{ZZZX_2006} & Bell state & 3 & Unidirectional & No & Impersonated fraudulent, direct measurement on channel particles,
attack on two-way channel\tabularnewline
 &  &  &  &  & \tabularnewline
Yuan et al. \cite{YLP+14} & Single photon & 1 & Unidirectional & No & Intercept-resend, measure-resend, entangle-measure on two-way channel\tabularnewline
 &  &  &  &  & \tabularnewline
Hong et al. \cite{HCJ+17} & Single photon & 1 & Unidirectional & No & Impersonation, measure resend, entangle-measure\tabularnewline
 &  &  &  &  & \tabularnewline
Kang et al. \cite{KHHYHM_2018,KHH+20} & GHZ-like & 6 & Bidirectional & Untrusted & $-$\tabularnewline
 &  &  &  &  & \tabularnewline
Zhang et al. \cite{ZCSL_2020} & Bell state & 6 & Bidirectional & Semi-honest & Impersonation, entangle and measure, intercept-resend, third party's\tabularnewline
 &  &  &  &  & \tabularnewline
Jiang et al. \cite{JZH21} & Bell state & $18,35$ & Bidirectional & No & Impersonation, intercept-measure-resend, entangle-measure\tabularnewline
 &  &  &  &  & \tabularnewline
Wu et al. \cite{WCG+21} & Bell state, GHZ state & 6 & Unidirectional & Semi-honest & External, dishonest participants\textquoteright , third party\textquoteright s,
impersonation\tabularnewline
 &  &  &  &  & \tabularnewline
Li et al. \cite{LZZ+22} & GHZ state & $10,3$ & Bidirectional & Trusted & Impersonation, entangle-measure, intercept-measure-resend, external\tabularnewline
 &  &  &  &  & \tabularnewline
Our protocol & Bell state & 4 & Bidirectional & Untrusted & Impersonation, intercept-resend, impersonated fraudulent/entangle-measure\tabularnewline
 &  &  &  &  & \tabularnewline
\hline 
\end{tabular*}
\par\end{centering}
\caption{\label{tab:Comparison}Detailed comparison with various previous QIA
protocols. The column ``Minimum secret key required'' contains multiple
values corresponding to different protocols within the same paper.}
\end{table}

\section{Conclusion\label{sec:VI}}

In an earlier work of the present authors \cite{DP22}, QIA schemes
were categorized based on the intrinsic quantum cryptographic tasks
involved in their design. It was observed that numerous existing schemes
for secure direct quantum communication and other cryptographic tasks
had been adapted to create schemes for QIA. However, the potential
for modifying all types of quantum cryptographic schemes for this
purpose remains largely unexplored. For instance, while some schemes
for QSDC and DSQC had been adapted for QIA, the possibility of modifying
their controlled versions or utilizing device-independent schemes
for QSDC, and CDSQC had not been investigated. To address this gap,
we present a novel QIA scheme by adapting a CDSQC scheme. In our protocol,
legitimate parties use Bell states as the quantum resource and Pauli
operations. With the help of an untrusted intermediary, Charlie, they
authenticate each other simultaneously. We compared our protocol with
previous protocols, demonstrating that our scheme achieves bidirectional
authentication using fewer quantum resources than some other protocols
with the help of an untrusted third party.

The present work, along with references \cite{DP22,DP23}, suggests
the potential for developing a wide range of new QIA schemes. A comprehensive
exploration in this direction can be highly valuable for identifying
the most efficient scheme for QIA that can be implemented using current
technology. Our proposed protocol for QIA is an entangled-state-based
scheme for QIA that utilizes Bell states. Inspired by the principles
of secure direct quantum communication, it enables Alice to securely
transmit information to Bob when controller Charlie authorizes it,
without the need for pre-generated keys. Although the protocol relies
solely on Bell states, its realization would necessitate quantum memory,
which is not yet commercially available. While this represents a limitation,
it is not unique to our protocol, but this limitation is common to
many existing protocols for QIA, QSDC, DSQC and quantum dialogue.
Given the growing interest in research and technology development
related to quantum memory, along with recent proposals for constructing
quantum memory, it is anticipated that quantum memory will become
available in the near future \cite{LLY+24}. In the interim period,
a delay mechanism can serve as a substitute for quantum memory in
implementing our proposed QIA protocol, thereby enhancing the resilience
of protocols for secure quantum communication. Additionally,
our work could be extended in the future with modifications that provide
resistance to collective noise \cite{TFI+16,WCS22,LWZ+16}.

\section*{Availability of data and materials}

No additional data is needed for this work.

\section*{Competing interests}

The authors declare that they have no competing interests.

\bibliographystyle{apsrev4-2}
\bibliography{MPLA_QIA}

\begin{thebibliography}{65}%
\makeatletter
\providecommand \@ifxundefined [1]{%
 \@ifx{#1\undefined}
}%
\providecommand \@ifnum [1]{%
 \ifnum #1\expandafter \@firstoftwo
 \else \expandafter \@secondoftwo
 \fi
}%
\providecommand \@ifx [1]{%
 \ifx #1\expandafter \@firstoftwo
 \else \expandafter \@secondoftwo
 \fi
}%
\providecommand \natexlab [1]{#1}%
\providecommand \enquote  [1]{``#1''}%
\providecommand \bibnamefont  [1]{#1}%
\providecommand \bibfnamefont [1]{#1}%
\providecommand \citenamefont [1]{#1}%
\providecommand \href@noop [0]{\@secondoftwo}%
\providecommand \href [0]{\begingroup \@sanitize@url \@href}%
\providecommand \@href[1]{\@@startlink{#1}\@@href}%
\providecommand \@@href[1]{\endgroup#1\@@endlink}%
\providecommand \@sanitize@url [0]{\catcode `\\12\catcode `\$12\catcode
  `\&12\catcode `\#12\catcode `\^12\catcode `\_12\catcode `\%12\relax}%
\providecommand \@@startlink[1]{}%
\providecommand \@@endlink[0]{}%
\providecommand \url  [0]{\begingroup\@sanitize@url \@url }%
\providecommand \@url [1]{\endgroup\@href {#1}{\urlprefix }}%
\providecommand \urlprefix  [0]{URL }%
\providecommand \Eprint [0]{\href }%
\providecommand \doibase [0]{https://doi.org/}%
\providecommand \selectlanguage [0]{\@gobble}%
\providecommand \bibinfo  [0]{\@secondoftwo}%
\providecommand \bibfield  [0]{\@secondoftwo}%
\providecommand \translation [1]{[#1]}%
\providecommand \BibitemOpen [0]{}%
\providecommand \bibitemStop [0]{}%
\providecommand \bibitemNoStop [0]{.\EOS\space}%
\providecommand \EOS [0]{\spacefactor3000\relax}%
\providecommand \BibitemShut  [1]{\csname bibitem#1\endcsname}%
\let\auto@bib@innerbib\@empty
\bibitem [{\citenamefont {Bennett}\ and\ \citenamefont
  {Brassard}(1984)}]{BB84}%
  \BibitemOpen
  \bibfield  {author} {\bibinfo {author} {\bibfnamefont {C.~H.}\ \bibnamefont
  {Bennett}}\ and\ \bibinfo {author} {\bibfnamefont {G.}~\bibnamefont
  {Brassard}},\ }\href@noop {} {\bibinfo {title} {Quantum cryptography:
  Public-key distribution and coin tossing, {in Proc. IEEE Int. Conf. on
  Computers, Systems, and Signal Processing (Bangalore, India, 1984), pp.
  175-179.}}} (\bibinfo {year} {1984})\BibitemShut {NoStop}%
\bibitem [{\citenamefont {Ekert}(1991)}]{E91}%
  \BibitemOpen
  \bibfield  {author} {\bibinfo {author} {\bibfnamefont {A.~K.}\ \bibnamefont
  {Ekert}},\ }\href@noop {} {\bibfield  {journal} {\bibinfo  {journal}
  {Physical Review Letters}\ }\textbf {\bibinfo {volume} {67}},\ \bibinfo
  {pages} {661} (\bibinfo {year} {1991})}\BibitemShut {NoStop}%
\bibitem [{\citenamefont {Bennett}(1992)}]{B92}%
  \BibitemOpen
  \bibfield  {author} {\bibinfo {author} {\bibfnamefont {C.~H.}\ \bibnamefont
  {Bennett}},\ }\href@noop {} {\bibfield  {journal} {\bibinfo  {journal}
  {Physical Review Letters}\ }\textbf {\bibinfo {volume} {68}},\ \bibinfo
  {pages} {3121} (\bibinfo {year} {1992})}\BibitemShut {NoStop}%
\bibitem [{\citenamefont {Wang}\ \emph {et~al.}(2021)\citenamefont {Wang},
  \citenamefont {Sun}, \citenamefont {Liu}, \citenamefont {Wang}, \citenamefont
  {Kan}, \citenamefont {Dong},\ and\ \citenamefont {Zhao}}]{WSL+21}%
  \BibitemOpen
  \bibfield  {author} {\bibinfo {author} {\bibfnamefont {X.-f.}\ \bibnamefont
  {Wang}}, \bibinfo {author} {\bibfnamefont {X.-j.}\ \bibnamefont {Sun}},
  \bibinfo {author} {\bibfnamefont {Y.-x.}\ \bibnamefont {Liu}}, \bibinfo
  {author} {\bibfnamefont {W.}~\bibnamefont {Wang}}, \bibinfo {author}
  {\bibfnamefont {B.-x.}\ \bibnamefont {Kan}}, \bibinfo {author} {\bibfnamefont
  {P.}~\bibnamefont {Dong}},\ and\ \bibinfo {author} {\bibfnamefont {L.-l.}\
  \bibnamefont {Zhao}},\ }\href@noop {} {\bibfield  {journal} {\bibinfo
  {journal} {Quantum Engineering}\ }\textbf {\bibinfo {volume} {3}},\ \bibinfo
  {pages} {e73} (\bibinfo {year} {2021})}\BibitemShut {NoStop}%
\bibitem [{\citenamefont {She}\ and\ \citenamefont {Zhang}(2022)}]{SZ22}%
  \BibitemOpen
  \bibfield  {author} {\bibinfo {author} {\bibfnamefont {L.-G.}\ \bibnamefont
  {She}}\ and\ \bibinfo {author} {\bibfnamefont {C.-M.}\ \bibnamefont
  {Zhang}},\ }\href@noop {} {\bibfield  {journal} {\bibinfo  {journal} {Quantum
  Information Processing}\ }\textbf {\bibinfo {volume} {21}},\ \bibinfo {pages}
  {161} (\bibinfo {year} {2022})}\BibitemShut {NoStop}%
\bibitem [{\citenamefont {Dutta}\ and\ \citenamefont
  {Pathak}(2022{\natexlab{a}})}]{DP+23}%
  \BibitemOpen
  \bibfield  {author} {\bibinfo {author} {\bibfnamefont {A.}~\bibnamefont
  {Dutta}}\ and\ \bibinfo {author} {\bibfnamefont {A.}~\bibnamefont {Pathak}},\
  }\href@noop {} {\bibfield  {journal} {\bibinfo  {journal} {arXiv preprint
  arXiv:2212.13089}\ } (\bibinfo {year} {2022}{\natexlab{a}})}\BibitemShut
  {NoStop}%
\bibitem [{\citenamefont {Dutta}\ \emph {et~al.}(2024)\citenamefont {Dutta},
  \citenamefont {Muskan}, \citenamefont {Banerjee},\ and\ \citenamefont
  {Pathak}}]{DMB+24}%
  \BibitemOpen
  \bibfield  {author} {\bibinfo {author} {\bibfnamefont {A.}~\bibnamefont
  {Dutta}}, \bibinfo {author} {\bibnamefont {Muskan}}, \bibinfo {author}
  {\bibfnamefont {S.}~\bibnamefont {Banerjee}},\ and\ \bibinfo {author}
  {\bibfnamefont {A.}~\bibnamefont {Pathak}},\ }\href@noop {} {\bibfield
  {journal} {\bibinfo  {journal} {Advanced Quantum Technologies}\ ,\ \bibinfo
  {pages} {2400149}} (\bibinfo {year} {2024})}\BibitemShut {NoStop}%
\bibitem [{\citenamefont {Lo}\ \emph {et~al.}(2012)\citenamefont {Lo},
  \citenamefont {Curty},\ and\ \citenamefont {Qi}}]{LCQ12}%
  \BibitemOpen
  \bibfield  {author} {\bibinfo {author} {\bibfnamefont {H.-K.}\ \bibnamefont
  {Lo}}, \bibinfo {author} {\bibfnamefont {M.}~\bibnamefont {Curty}},\ and\
  \bibinfo {author} {\bibfnamefont {B.}~\bibnamefont {Qi}},\ }\href@noop {}
  {\bibfield  {journal} {\bibinfo  {journal} {Physical Review Letters}\
  }\textbf {\bibinfo {volume} {108}},\ \bibinfo {pages} {130503} (\bibinfo
  {year} {2012})}\BibitemShut {NoStop}%
\bibitem [{\citenamefont {Cao}\ \emph {et~al.}(2020)\citenamefont {Cao},
  \citenamefont {Li}, \citenamefont {Yang}, \citenamefont {Jiang},
  \citenamefont {Li}, \citenamefont {Hu}, \citenamefont {Abulizi},
  \citenamefont {Li}, \citenamefont {Zhang}, \citenamefont {Sun} \emph
  {et~al.}}]{CLY+20}%
  \BibitemOpen
  \bibfield  {author} {\bibinfo {author} {\bibfnamefont {Y.}~\bibnamefont
  {Cao}}, \bibinfo {author} {\bibfnamefont {Y.-H.}\ \bibnamefont {Li}},
  \bibinfo {author} {\bibfnamefont {K.-X.}\ \bibnamefont {Yang}}, \bibinfo
  {author} {\bibfnamefont {Y.-F.}\ \bibnamefont {Jiang}}, \bibinfo {author}
  {\bibfnamefont {S.-L.}\ \bibnamefont {Li}}, \bibinfo {author} {\bibfnamefont
  {X.-L.}\ \bibnamefont {Hu}}, \bibinfo {author} {\bibfnamefont
  {M.}~\bibnamefont {Abulizi}}, \bibinfo {author} {\bibfnamefont {C.-L.}\
  \bibnamefont {Li}}, \bibinfo {author} {\bibfnamefont {W.}~\bibnamefont
  {Zhang}}, \bibinfo {author} {\bibfnamefont {Q.-C.}\ \bibnamefont {Sun}},
  \emph {et~al.},\ }\href@noop {} {\bibfield  {journal} {\bibinfo  {journal}
  {Physical Review Letters}\ }\textbf {\bibinfo {volume} {125}},\ \bibinfo
  {pages} {260503} (\bibinfo {year} {2020})}\BibitemShut {NoStop}%
\bibitem [{\citenamefont {Jouguet}\ \emph {et~al.}(2013)\citenamefont
  {Jouguet}, \citenamefont {Kunz-Jacques}, \citenamefont {Leverrier},
  \citenamefont {Grangier},\ and\ \citenamefont {Diamanti}}]{JJL+13}%
  \BibitemOpen
  \bibfield  {author} {\bibinfo {author} {\bibfnamefont {P.}~\bibnamefont
  {Jouguet}}, \bibinfo {author} {\bibfnamefont {S.}~\bibnamefont
  {Kunz-Jacques}}, \bibinfo {author} {\bibfnamefont {A.}~\bibnamefont
  {Leverrier}}, \bibinfo {author} {\bibfnamefont {P.}~\bibnamefont
  {Grangier}},\ and\ \bibinfo {author} {\bibfnamefont {E.}~\bibnamefont
  {Diamanti}},\ }\href@noop {} {\bibfield  {journal} {\bibinfo  {journal}
  {Nature Photonics}\ }\textbf {\bibinfo {volume} {7}},\ \bibinfo {pages} {378}
  (\bibinfo {year} {2013})}\BibitemShut {NoStop}%
\bibitem [{\citenamefont {Hu}\ \emph {et~al.}(2020)\citenamefont {Hu},
  \citenamefont {Al-Amri}, \citenamefont {Liao},\ and\ \citenamefont
  {Zubairy}}]{HAL+20}%
  \BibitemOpen
  \bibfield  {author} {\bibinfo {author} {\bibfnamefont {L.}~\bibnamefont
  {Hu}}, \bibinfo {author} {\bibfnamefont {M.}~\bibnamefont {Al-Amri}},
  \bibinfo {author} {\bibfnamefont {Z.}~\bibnamefont {Liao}},\ and\ \bibinfo
  {author} {\bibfnamefont {M.}~\bibnamefont {Zubairy}},\ }\href@noop {}
  {\bibfield  {journal} {\bibinfo  {journal} {Physical Review A}\ }\textbf
  {\bibinfo {volume} {102}},\ \bibinfo {pages} {012608} (\bibinfo {year}
  {2020})}\BibitemShut {NoStop}%
\bibitem [{\citenamefont {Wang}\ \emph
  {et~al.}(2022{\natexlab{a}})\citenamefont {Wang}, \citenamefont {Yin},
  \citenamefont {He}, \citenamefont {Chen}, \citenamefont {Wang}, \citenamefont
  {Ye}, \citenamefont {Zhou}, \citenamefont {Fan-Yuan}, \citenamefont {Wang},
  \citenamefont {Chen} \emph {et~al.}}]{WYH+22}%
  \BibitemOpen
  \bibfield  {author} {\bibinfo {author} {\bibfnamefont {S.}~\bibnamefont
  {Wang}}, \bibinfo {author} {\bibfnamefont {Z.-Q.}\ \bibnamefont {Yin}},
  \bibinfo {author} {\bibfnamefont {D.-Y.}\ \bibnamefont {He}}, \bibinfo
  {author} {\bibfnamefont {W.}~\bibnamefont {Chen}}, \bibinfo {author}
  {\bibfnamefont {R.-Q.}\ \bibnamefont {Wang}}, \bibinfo {author}
  {\bibfnamefont {P.}~\bibnamefont {Ye}}, \bibinfo {author} {\bibfnamefont
  {Y.}~\bibnamefont {Zhou}}, \bibinfo {author} {\bibfnamefont {G.-J.}\
  \bibnamefont {Fan-Yuan}}, \bibinfo {author} {\bibfnamefont {F.-X.}\
  \bibnamefont {Wang}}, \bibinfo {author} {\bibfnamefont {W.}~\bibnamefont
  {Chen}}, \emph {et~al.},\ }\href@noop {} {\bibfield  {journal} {\bibinfo
  {journal} {Nature Photonics}\ }\textbf {\bibinfo {volume} {16}},\ \bibinfo
  {pages} {154} (\bibinfo {year} {2022}{\natexlab{a}})}\BibitemShut {NoStop}%
\bibitem [{\citenamefont {Wegman}\ and\ \citenamefont {Carter}(1981)}]{WC81}%
  \BibitemOpen
  \bibfield  {author} {\bibinfo {author} {\bibfnamefont {M.~N.}\ \bibnamefont
  {Wegman}}\ and\ \bibinfo {author} {\bibfnamefont {J.~L.}\ \bibnamefont
  {Carter}},\ }\href@noop {} {\bibfield  {journal} {\bibinfo  {journal}
  {Journal of Computer and System Sciences}\ }\textbf {\bibinfo {volume}
  {22}},\ \bibinfo {pages} {265} (\bibinfo {year} {1981})}\BibitemShut
  {NoStop}%
\bibitem [{\citenamefont {Cr{\'e}peau}\ and\ \citenamefont
  {Salvail}(1995)}]{CS_1995}%
  \BibitemOpen
  \bibfield  {author} {\bibinfo {author} {\bibfnamefont {C.}~\bibnamefont
  {Cr{\'e}peau}}\ and\ \bibinfo {author} {\bibfnamefont {L.}~\bibnamefont
  {Salvail}},\ }in\ \href@noop {} {\emph {\bibinfo {booktitle} {International
  Conference on the Theory and Applications of Cryptographic Techniques}}}\
  (\bibinfo {organization} {Springer},\ \bibinfo {year} {1995})\ pp.\ \bibinfo
  {pages} {133--146}\BibitemShut {NoStop}%
\bibitem [{\citenamefont {Li}\ and\ \citenamefont {Barnum}(2004)}]{LB_2004}%
  \BibitemOpen
  \bibfield  {author} {\bibinfo {author} {\bibfnamefont {X.}~\bibnamefont
  {Li}}\ and\ \bibinfo {author} {\bibfnamefont {H.}~\bibnamefont {Barnum}},\
  }\href@noop {} {\bibfield  {journal} {\bibinfo  {journal} {International
  Journal of Foundations of Computer Science}\ }\textbf {\bibinfo {volume}
  {15}},\ \bibinfo {pages} {609} (\bibinfo {year} {2004})}\BibitemShut
  {NoStop}%
\bibitem [{\citenamefont {WANG}\ \emph {et~al.}(2006)\citenamefont {WANG},
  \citenamefont {ZHANG},\ and\ \citenamefont {TANG}}]{WZT_2006}%
  \BibitemOpen
  \bibfield  {author} {\bibinfo {author} {\bibfnamefont {J.}~\bibnamefont
  {WANG}}, \bibinfo {author} {\bibfnamefont {Q.}~\bibnamefont {ZHANG}},\ and\
  \bibinfo {author} {\bibfnamefont {C.-J.}\ \bibnamefont {TANG}},\ }\href@noop
  {} {\bibfield  {journal} {\bibinfo  {journal} {Chinese Physics Letters}\
  }\textbf {\bibinfo {volume} {23}},\ \bibinfo {pages} {2360} (\bibinfo {year}
  {2006})}\BibitemShut {NoStop}%
\bibitem [{\citenamefont {Zhang}\ \emph {et~al.}(2006)\citenamefont {Zhang},
  \citenamefont {Zeng}, \citenamefont {Zhou},\ and\ \citenamefont
  {Xiong}}]{ZZZX_2006}%
  \BibitemOpen
  \bibfield  {author} {\bibinfo {author} {\bibfnamefont {Z.}~\bibnamefont
  {Zhang}}, \bibinfo {author} {\bibfnamefont {G.}~\bibnamefont {Zeng}},
  \bibinfo {author} {\bibfnamefont {N.}~\bibnamefont {Zhou}},\ and\ \bibinfo
  {author} {\bibfnamefont {J.}~\bibnamefont {Xiong}},\ }\href@noop {}
  {\bibfield  {journal} {\bibinfo  {journal} {Physics Letters A}\ }\textbf
  {\bibinfo {volume} {356}},\ \bibinfo {pages} {199} (\bibinfo {year}
  {2006})}\BibitemShut {NoStop}%
\bibitem [{\citenamefont {Zhang}\ \emph {et~al.}(2020)\citenamefont {Zhang},
  \citenamefont {Chen}, \citenamefont {Shi},\ and\ \citenamefont
  {Liang}}]{ZCSL_2020}%
  \BibitemOpen
  \bibfield  {author} {\bibinfo {author} {\bibfnamefont {S.}~\bibnamefont
  {Zhang}}, \bibinfo {author} {\bibfnamefont {Z.-K.}\ \bibnamefont {Chen}},
  \bibinfo {author} {\bibfnamefont {R.-H.}\ \bibnamefont {Shi}},\ and\ \bibinfo
  {author} {\bibfnamefont {F.-Y.}\ \bibnamefont {Liang}},\ }\href@noop {}
  {\bibfield  {journal} {\bibinfo  {journal} {International Journal of
  Theoretical Physics}\ }\textbf {\bibinfo {volume} {59}},\ \bibinfo {pages}
  {236} (\bibinfo {year} {2020})}\BibitemShut {NoStop}%
\bibitem [{\citenamefont {Kang}\ \emph {et~al.}(2018)\citenamefont {Kang},
  \citenamefont {Heo}, \citenamefont {Hong}, \citenamefont {Yang},
  \citenamefont {Han},\ and\ \citenamefont {Moon}}]{KHHYHM_2018}%
  \BibitemOpen
  \bibfield  {author} {\bibinfo {author} {\bibfnamefont {M.-S.}\ \bibnamefont
  {Kang}}, \bibinfo {author} {\bibfnamefont {J.}~\bibnamefont {Heo}}, \bibinfo
  {author} {\bibfnamefont {C.-H.}\ \bibnamefont {Hong}}, \bibinfo {author}
  {\bibfnamefont {H.-J.}\ \bibnamefont {Yang}}, \bibinfo {author}
  {\bibfnamefont {S.-W.}\ \bibnamefont {Han}},\ and\ \bibinfo {author}
  {\bibfnamefont {S.}~\bibnamefont {Moon}},\ }\href@noop {} {\bibfield
  {journal} {\bibinfo  {journal} {Quantum Information Processing}\ }\textbf
  {\bibinfo {volume} {17}},\ \bibinfo {pages} {159} (\bibinfo {year}
  {2018})}\BibitemShut {NoStop}%
\bibitem [{\citenamefont {Chang}\ \emph {et~al.}(2014)\citenamefont {Chang},
  \citenamefont {Xu}, \citenamefont {Zhang},\ and\ \citenamefont
  {Yan}}]{CXZY_2014}%
  \BibitemOpen
  \bibfield  {author} {\bibinfo {author} {\bibfnamefont {Y.}~\bibnamefont
  {Chang}}, \bibinfo {author} {\bibfnamefont {C.}~\bibnamefont {Xu}}, \bibinfo
  {author} {\bibfnamefont {S.}~\bibnamefont {Zhang}},\ and\ \bibinfo {author}
  {\bibfnamefont {L.}~\bibnamefont {Yan}},\ }\href@noop {} {\bibfield
  {journal} {\bibinfo  {journal} {Chinese Science Bulletin}\ }\textbf {\bibinfo
  {volume} {59}},\ \bibinfo {pages} {2541} (\bibinfo {year}
  {2014})}\BibitemShut {NoStop}%
\bibitem [{\citenamefont {Mihara}(2002)}]{T.Mihara_2002}%
  \BibitemOpen
  \bibfield  {author} {\bibinfo {author} {\bibfnamefont {T.}~\bibnamefont
  {Mihara}},\ }\href@noop {} {\bibfield  {journal} {\bibinfo  {journal}
  {Physical Review A}\ }\textbf {\bibinfo {volume} {65}},\ \bibinfo {pages}
  {052326} (\bibinfo {year} {2002})}\BibitemShut {NoStop}%
\bibitem [{\citenamefont {Dutta}\ and\ \citenamefont
  {Pathak}(2022{\natexlab{b}})}]{DP22}%
  \BibitemOpen
  \bibfield  {author} {\bibinfo {author} {\bibfnamefont {A.}~\bibnamefont
  {Dutta}}\ and\ \bibinfo {author} {\bibfnamefont {A.}~\bibnamefont {Pathak}},\
  }\href@noop {} {\bibfield  {journal} {\bibinfo  {journal} {Quantum
  Information Processing}\ }\textbf {\bibinfo {volume} {21}},\ \bibinfo {pages}
  {369} (\bibinfo {year} {2022}{\natexlab{b}})}\BibitemShut {NoStop}%
\bibitem [{\citenamefont {Dutta}\ and\ \citenamefont {Pathak}(2023)}]{DP23}%
  \BibitemOpen
  \bibfield  {author} {\bibinfo {author} {\bibfnamefont {A.}~\bibnamefont
  {Dutta}}\ and\ \bibinfo {author} {\bibfnamefont {A.}~\bibnamefont {Pathak}},\
  }\href@noop {} {\bibfield  {journal} {\bibinfo  {journal} {Quantum
  Information Processing}\ }\textbf {\bibinfo {volume} {22}},\ \bibinfo {pages}
  {13} (\bibinfo {year} {2023})}\BibitemShut {NoStop}%
\bibitem [{\citenamefont {Jian}\ \emph {et~al.}(2023)\citenamefont {Jian},
  \citenamefont {Wang}, \citenamefont {Chen}, \citenamefont {Zhou},\ and\
  \citenamefont {Liu}}]{JWC+23}%
  \BibitemOpen
  \bibfield  {author} {\bibinfo {author} {\bibfnamefont {L.}~\bibnamefont
  {Jian}}, \bibinfo {author} {\bibfnamefont {Y.}~\bibnamefont {Wang}}, \bibinfo
  {author} {\bibfnamefont {G.}~\bibnamefont {Chen}}, \bibinfo {author}
  {\bibfnamefont {Y.}~\bibnamefont {Zhou}},\ and\ \bibinfo {author}
  {\bibfnamefont {S.}~\bibnamefont {Liu}},\ }\href@noop {} {\bibfield
  {journal} {\bibinfo  {journal} {Journal of Physics B: Atomic, Molecular and
  Optical Physics}\ }\textbf {\bibinfo {volume} {56}},\ \bibinfo {pages}
  {075502} (\bibinfo {year} {2023})}\BibitemShut {NoStop}%
\bibitem [{\citenamefont {Li}\ \emph {et~al.}(2022)\citenamefont {Li},
  \citenamefont {Zhang}, \citenamefont {Zhang},\ and\ \citenamefont
  {Zhao}}]{LZZ+22}%
  \BibitemOpen
  \bibfield  {author} {\bibinfo {author} {\bibfnamefont {X.}~\bibnamefont
  {Li}}, \bibinfo {author} {\bibfnamefont {K.}~\bibnamefont {Zhang}}, \bibinfo
  {author} {\bibfnamefont {L.}~\bibnamefont {Zhang}},\ and\ \bibinfo {author}
  {\bibfnamefont {X.}~\bibnamefont {Zhao}},\ }\href@noop {} {\bibfield
  {journal} {\bibinfo  {journal} {Entropy}\ }\textbf {\bibinfo {volume} {24}},\
  \bibinfo {pages} {483} (\bibinfo {year} {2022})}\BibitemShut {NoStop}%
\bibitem [{\citenamefont {Faleiro}\ and\ \citenamefont
  {Goul{\~a}o}(2021)}]{FG21}%
  \BibitemOpen
  \bibfield  {author} {\bibinfo {author} {\bibfnamefont {R.}~\bibnamefont
  {Faleiro}}\ and\ \bibinfo {author} {\bibfnamefont {M.}~\bibnamefont
  {Goul{\~a}o}},\ }\href@noop {} {\bibfield  {journal} {\bibinfo  {journal}
  {Physical Review A}\ }\textbf {\bibinfo {volume} {103}},\ \bibinfo {pages}
  {022430} (\bibinfo {year} {2021})}\BibitemShut {NoStop}%
\bibitem [{\citenamefont {Chen}\ \emph {et~al.}(2023)\citenamefont {Chen},
  \citenamefont {Wang}, \citenamefont {Jian}, \citenamefont {Zhou},
  \citenamefont {Liu}, \citenamefont {Luo},\ and\ \citenamefont
  {Yang}}]{CWJ+23}%
  \BibitemOpen
  \bibfield  {author} {\bibinfo {author} {\bibfnamefont {G.}~\bibnamefont
  {Chen}}, \bibinfo {author} {\bibfnamefont {Y.}~\bibnamefont {Wang}}, \bibinfo
  {author} {\bibfnamefont {L.}~\bibnamefont {Jian}}, \bibinfo {author}
  {\bibfnamefont {Y.}~\bibnamefont {Zhou}}, \bibinfo {author} {\bibfnamefont
  {S.}~\bibnamefont {Liu}}, \bibinfo {author} {\bibfnamefont {J.}~\bibnamefont
  {Luo}},\ and\ \bibinfo {author} {\bibfnamefont {K.}~\bibnamefont {Yang}},\
  }\href@noop {} {\bibfield  {journal} {\bibinfo  {journal} {Journal of Applied
  Physics}\ }\textbf {\bibinfo {volume} {133}} (\bibinfo {year}
  {2023})}\BibitemShut {NoStop}%
\bibitem [{\citenamefont {Lo}\ and\ \citenamefont {Chau}(1997)}]{LC_1997}%
  \BibitemOpen
  \bibfield  {author} {\bibinfo {author} {\bibfnamefont {H.-K.}\ \bibnamefont
  {Lo}}\ and\ \bibinfo {author} {\bibfnamefont {H.~F.}\ \bibnamefont {Chau}},\
  }\href@noop {} {\bibfield  {journal} {\bibinfo  {journal} {Physical Review
  Letters}\ }\textbf {\bibinfo {volume} {78}},\ \bibinfo {pages} {3410}
  (\bibinfo {year} {1997})}\BibitemShut {NoStop}%
\bibitem [{\citenamefont {Zeng}\ and\ \citenamefont {Wang}(1998)}]{ZW_1998}%
  \BibitemOpen
  \bibfield  {author} {\bibinfo {author} {\bibfnamefont {G.}~\bibnamefont
  {Zeng}}\ and\ \bibinfo {author} {\bibfnamefont {X.}~\bibnamefont {Wang}},\
  }\href@noop {} {\bibfield  {journal} {\bibinfo  {journal} {arXiv preprint
  quant-ph/9812022}\ } (\bibinfo {year} {1998})}\BibitemShut {NoStop}%
\bibitem [{\citenamefont {Du{\v{s}}ek}\ \emph {et~al.}(1999)\citenamefont
  {Du{\v{s}}ek}, \citenamefont {Haderka}, \citenamefont {Hendrych},\ and\
  \citenamefont {My{\v{s}}ka}}]{DHHM_1999}%
  \BibitemOpen
  \bibfield  {author} {\bibinfo {author} {\bibfnamefont {M.}~\bibnamefont
  {Du{\v{s}}ek}}, \bibinfo {author} {\bibfnamefont {O.}~\bibnamefont
  {Haderka}}, \bibinfo {author} {\bibfnamefont {M.}~\bibnamefont {Hendrych}},\
  and\ \bibinfo {author} {\bibfnamefont {R.}~\bibnamefont {My{\v{s}}ka}},\
  }\href@noop {} {\bibfield  {journal} {\bibinfo  {journal} {Physical Review
  A}\ }\textbf {\bibinfo {volume} {60}},\ \bibinfo {pages} {149} (\bibinfo
  {year} {1999})}\BibitemShut {NoStop}%
\bibitem [{\citenamefont {Long}\ and\ \citenamefont {Liu}(2002)}]{LL02}%
  \BibitemOpen
  \bibfield  {author} {\bibinfo {author} {\bibfnamefont {G.-L.}\ \bibnamefont
  {Long}}\ and\ \bibinfo {author} {\bibfnamefont {X.-S.}\ \bibnamefont {Liu}},\
  }\href@noop {} {\bibfield  {journal} {\bibinfo  {journal} {Physical Review
  A}\ }\textbf {\bibinfo {volume} {65}},\ \bibinfo {pages} {032302} (\bibinfo
  {year} {2002})}\BibitemShut {NoStop}%
\bibitem [{\citenamefont {Lucamarini}\ and\ \citenamefont
  {Mancini}(2005)}]{LM_05}%
  \BibitemOpen
  \bibfield  {author} {\bibinfo {author} {\bibfnamefont {M.}~\bibnamefont
  {Lucamarini}}\ and\ \bibinfo {author} {\bibfnamefont {S.}~\bibnamefont
  {Mancini}},\ }\href@noop {} {\bibfield  {journal} {\bibinfo  {journal}
  {Physical Review Letters}\ }\textbf {\bibinfo {volume} {94}},\ \bibinfo
  {pages} {140501} (\bibinfo {year} {2005})}\BibitemShut {NoStop}%
\bibitem [{\citenamefont {Pathak}(2013)}]{P13}%
  \BibitemOpen
  \bibfield  {author} {\bibinfo {author} {\bibfnamefont {A.}~\bibnamefont
  {Pathak}},\ }\href@noop {} {\emph {\bibinfo {title} {Elements of quantum
  computation and quantum communication}}}\ (\bibinfo  {publisher} {CRC Press
  Boca Raton},\ \bibinfo {year} {2013})\BibitemShut {NoStop}%
\bibitem [{\citenamefont {Srikara}\ \emph {et~al.}(2020)\citenamefont
  {Srikara}, \citenamefont {Thapliyal},\ and\ \citenamefont {Pathak}}]{STP20}%
  \BibitemOpen
  \bibfield  {author} {\bibinfo {author} {\bibfnamefont {S.}~\bibnamefont
  {Srikara}}, \bibinfo {author} {\bibfnamefont {K.}~\bibnamefont {Thapliyal}},\
  and\ \bibinfo {author} {\bibfnamefont {A.}~\bibnamefont {Pathak}},\
  }\href@noop {} {\bibfield  {journal} {\bibinfo  {journal} {Quantum
  Information Processing}\ }\textbf {\bibinfo {volume} {19}},\ \bibinfo {pages}
  {132} (\bibinfo {year} {2020})}\BibitemShut {NoStop}%
\bibitem [{\citenamefont {Shukla}\ \emph {et~al.}(2017)\citenamefont {Shukla},
  \citenamefont {Thapliyal},\ and\ \citenamefont {Pathak}}]{STP17}%
  \BibitemOpen
  \bibfield  {author} {\bibinfo {author} {\bibfnamefont {C.}~\bibnamefont
  {Shukla}}, \bibinfo {author} {\bibfnamefont {K.}~\bibnamefont {Thapliyal}},\
  and\ \bibinfo {author} {\bibfnamefont {A.}~\bibnamefont {Pathak}},\
  }\href@noop {} {\bibfield  {journal} {\bibinfo  {journal} {Quantum
  Information Processing}\ }\textbf {\bibinfo {volume} {16}},\ \bibinfo {pages}
  {295} (\bibinfo {year} {2017})}\BibitemShut {NoStop}%
\bibitem [{\citenamefont {Yadav}\ \emph {et~al.}(2014)\citenamefont {Yadav},
  \citenamefont {Srikanth},\ and\ \citenamefont {Pathak}}]{YSP14}%
  \BibitemOpen
  \bibfield  {author} {\bibinfo {author} {\bibfnamefont {P.}~\bibnamefont
  {Yadav}}, \bibinfo {author} {\bibfnamefont {R.}~\bibnamefont {Srikanth}},\
  and\ \bibinfo {author} {\bibfnamefont {A.}~\bibnamefont {Pathak}},\
  }\href@noop {} {\bibfield  {journal} {\bibinfo  {journal} {Quantum
  Information Processing}\ }\textbf {\bibinfo {volume} {13}},\ \bibinfo {pages}
  {2731} (\bibinfo {year} {2014})}\BibitemShut {NoStop}%
\bibitem [{\citenamefont {Banerjee}\ and\ \citenamefont {Pathak}(2012)}]{BP12}%
  \BibitemOpen
  \bibfield  {author} {\bibinfo {author} {\bibfnamefont {A.}~\bibnamefont
  {Banerjee}}\ and\ \bibinfo {author} {\bibfnamefont {A.}~\bibnamefont
  {Pathak}},\ }\href@noop {} {\bibfield  {journal} {\bibinfo  {journal}
  {Physics Letters A}\ }\textbf {\bibinfo {volume} {376}},\ \bibinfo {pages}
  {2944} (\bibinfo {year} {2012})}\BibitemShut {NoStop}%
\bibitem [{\citenamefont {Deng}\ \emph {et~al.}(2003)\citenamefont {Deng},
  \citenamefont {Long},\ and\ \citenamefont {Liu}}]{DLL03}%
  \BibitemOpen
  \bibfield  {author} {\bibinfo {author} {\bibfnamefont {F.-G.}\ \bibnamefont
  {Deng}}, \bibinfo {author} {\bibfnamefont {G.~L.}\ \bibnamefont {Long}},\
  and\ \bibinfo {author} {\bibfnamefont {X.-S.}\ \bibnamefont {Liu}},\
  }\href@noop {} {\bibfield  {journal} {\bibinfo  {journal} {Physical Review
  A}\ }\textbf {\bibinfo {volume} {68}},\ \bibinfo {pages} {042317} (\bibinfo
  {year} {2003})}\BibitemShut {NoStop}%
\bibitem [{\citenamefont {Dutta}\ and\ \citenamefont {Pathak}(2024)}]{DP24}%
  \BibitemOpen
  \bibfield  {author} {\bibinfo {author} {\bibfnamefont {A.}~\bibnamefont
  {Dutta}}\ and\ \bibinfo {author} {\bibfnamefont {A.}~\bibnamefont {Pathak}},\
  }\href@noop {} {\bibfield  {journal} {\bibinfo  {journal} {Physica Scripta}\
  }\textbf {\bibinfo {volume} {99}},\ \bibinfo {pages} {095106} (\bibinfo
  {year} {2024})}\BibitemShut {NoStop}%
\bibitem [{\citenamefont {Paparelle}\ \emph {et~al.}(2023)\citenamefont
  {Paparelle}, \citenamefont {Mousavi}, \citenamefont {Scazza}, \citenamefont
  {Paris}, \citenamefont {Bassi},\ and\ \citenamefont {Zavatta}}]{PMS+23}%
  \BibitemOpen
  \bibfield  {author} {\bibinfo {author} {\bibfnamefont {I.}~\bibnamefont
  {Paparelle}}, \bibinfo {author} {\bibfnamefont {F.}~\bibnamefont {Mousavi}},
  \bibinfo {author} {\bibfnamefont {F.}~\bibnamefont {Scazza}}, \bibinfo
  {author} {\bibfnamefont {M.}~\bibnamefont {Paris}}, \bibinfo {author}
  {\bibfnamefont {A.}~\bibnamefont {Bassi}},\ and\ \bibinfo {author}
  {\bibfnamefont {A.}~\bibnamefont {Zavatta}},\ }in\ \href@noop {} {\emph
  {\bibinfo {booktitle} {European Quantum Electronics Conference}}}\ (\bibinfo
  {organization} {Optica Publishing Group},\ \bibinfo {year} {2023})\ p.\
  \bibinfo {pages} {eb\_5\_4}\BibitemShut {NoStop}%
\bibitem [{\citenamefont {Bostr{\"o}m}\ and\ \citenamefont
  {Felbinger}(2002)}]{BF02}%
  \BibitemOpen
  \bibfield  {author} {\bibinfo {author} {\bibfnamefont {K.}~\bibnamefont
  {Bostr{\"o}m}}\ and\ \bibinfo {author} {\bibfnamefont {T.}~\bibnamefont
  {Felbinger}},\ }\href@noop {} {\bibfield  {journal} {\bibinfo  {journal}
  {Physical Review Letters}\ }\textbf {\bibinfo {volume} {89}},\ \bibinfo
  {pages} {187902} (\bibinfo {year} {2002})}\BibitemShut {NoStop}%
\bibitem [{\citenamefont {Yuan}\ \emph {et~al.}(2014)\citenamefont {Yuan},
  \citenamefont {Liu}, \citenamefont {Pan}, \citenamefont {Zhang},
  \citenamefont {Zhou},\ and\ \citenamefont {Zhang}}]{YLP+14}%
  \BibitemOpen
  \bibfield  {author} {\bibinfo {author} {\bibfnamefont {H.}~\bibnamefont
  {Yuan}}, \bibinfo {author} {\bibfnamefont {Y.-M.}\ \bibnamefont {Liu}},
  \bibinfo {author} {\bibfnamefont {G.-Z.}\ \bibnamefont {Pan}}, \bibinfo
  {author} {\bibfnamefont {G.}~\bibnamefont {Zhang}}, \bibinfo {author}
  {\bibfnamefont {J.}~\bibnamefont {Zhou}},\ and\ \bibinfo {author}
  {\bibfnamefont {Z.-J.}\ \bibnamefont {Zhang}},\ }\href@noop {} {\bibfield
  {journal} {\bibinfo  {journal} {Quantum Information Processing}\ }\textbf
  {\bibinfo {volume} {13}},\ \bibinfo {pages} {2535} (\bibinfo {year}
  {2014})}\BibitemShut {NoStop}%
\bibitem [{\citenamefont {Zhang}\ \emph {et~al.}(2000)\citenamefont {Zhang},
  \citenamefont {Li},\ and\ \citenamefont {Guo}}]{ZLG_2000}%
  \BibitemOpen
  \bibfield  {author} {\bibinfo {author} {\bibfnamefont {Y.-S.}\ \bibnamefont
  {Zhang}}, \bibinfo {author} {\bibfnamefont {C.-F.}\ \bibnamefont {Li}},\ and\
  \bibinfo {author} {\bibfnamefont {G.-C.}\ \bibnamefont {Guo}},\ }\href@noop
  {} {\bibfield  {journal} {\bibinfo  {journal} {arXiv preprint
  quant-ph/0008044}\ } (\bibinfo {year} {2000})}\BibitemShut {NoStop}%
\bibitem [{\citenamefont {Li}\ and\ \citenamefont {Chen}(2007)}]{LC_2007}%
  \BibitemOpen
  \bibfield  {author} {\bibinfo {author} {\bibfnamefont {X.}~\bibnamefont
  {Li}}\ and\ \bibinfo {author} {\bibfnamefont {L.}~\bibnamefont {Chen}},\ }in\
  \href@noop {} {\emph {\bibinfo {booktitle} {The First International Symposium
  on Data, Privacy, and E-Commerce (ISDPE 2007)}}}\ (\bibinfo {organization}
  {IEEE},\ \bibinfo {year} {2007})\ pp.\ \bibinfo {pages}
  {128--132}\BibitemShut {NoStop}%
\bibitem [{\citenamefont {Pathak}(2015)}]{P15}%
  \BibitemOpen
  \bibfield  {author} {\bibinfo {author} {\bibfnamefont {A.}~\bibnamefont
  {Pathak}},\ }\href@noop {} {\bibfield  {journal} {\bibinfo  {journal}
  {Quantum Information Processing}\ }\textbf {\bibinfo {volume} {14}},\
  \bibinfo {pages} {2195} (\bibinfo {year} {2015})}\BibitemShut {NoStop}%
\bibitem [{\citenamefont {Thapliyal}\ and\ \citenamefont
  {Pathak}(2015)}]{TP15}%
  \BibitemOpen
  \bibfield  {author} {\bibinfo {author} {\bibfnamefont {K.}~\bibnamefont
  {Thapliyal}}\ and\ \bibinfo {author} {\bibfnamefont {A.}~\bibnamefont
  {Pathak}},\ }\href@noop {} {\bibfield  {journal} {\bibinfo  {journal}
  {Quantum Information Processing}\ }\textbf {\bibinfo {volume} {14}},\
  \bibinfo {pages} {2599} (\bibinfo {year} {2015})}\BibitemShut {NoStop}%
\bibitem [{\citenamefont {Shukla}\ \emph {et~al.}(2012)\citenamefont {Shukla},
  \citenamefont {Pathak},\ and\ \citenamefont {Srikanth}}]{SPS12}%
  \BibitemOpen
  \bibfield  {author} {\bibinfo {author} {\bibfnamefont {C.}~\bibnamefont
  {Shukla}}, \bibinfo {author} {\bibfnamefont {A.}~\bibnamefont {Pathak}},\
  and\ \bibinfo {author} {\bibfnamefont {R.}~\bibnamefont {Srikanth}},\
  }\href@noop {} {\bibfield  {journal} {\bibinfo  {journal} {International
  Journal of Quantum Information}\ }\textbf {\bibinfo {volume} {10}},\ \bibinfo
  {pages} {1241009} (\bibinfo {year} {2012})}\BibitemShut {NoStop}%
\bibitem [{\citenamefont {Lo}\ \emph {et~al.}(2005)\citenamefont {Lo},
  \citenamefont {Ma},\ and\ \citenamefont {Chen}}]{LMC05}%
  \BibitemOpen
  \bibfield  {author} {\bibinfo {author} {\bibfnamefont {H.-K.}\ \bibnamefont
  {Lo}}, \bibinfo {author} {\bibfnamefont {X.}~\bibnamefont {Ma}},\ and\
  \bibinfo {author} {\bibfnamefont {K.}~\bibnamefont {Chen}},\ }\href@noop {}
  {\bibfield  {journal} {\bibinfo  {journal} {Physical Review Letters}\
  }\textbf {\bibinfo {volume} {94}},\ \bibinfo {pages} {230504} (\bibinfo
  {year} {2005})}\BibitemShut {NoStop}%
\bibitem [{\citenamefont {Holevo}(1973)}]{H_73}%
  \BibitemOpen
  \bibfield  {author} {\bibinfo {author} {\bibfnamefont {A.~S.}\ \bibnamefont
  {Holevo}},\ }\href@noop {} {\bibfield  {journal} {\bibinfo  {journal}
  {Problemy Peredachi Informatsii}\ }\textbf {\bibinfo {volume} {9}},\ \bibinfo
  {pages} {3} (\bibinfo {year} {1973})}\BibitemShut {NoStop}%
\bibitem [{\citenamefont {Christandl}\ \emph {et~al.}(2004)\citenamefont
  {Christandl}, \citenamefont {Renner},\ and\ \citenamefont {Ekert}}]{CRE_04}%
  \BibitemOpen
  \bibfield  {author} {\bibinfo {author} {\bibfnamefont {M.}~\bibnamefont
  {Christandl}}, \bibinfo {author} {\bibfnamefont {R.}~\bibnamefont {Renner}},\
  and\ \bibinfo {author} {\bibfnamefont {A.}~\bibnamefont {Ekert}},\
  }\href@noop {} {\bibfield  {journal} {\bibinfo  {journal} {arXiv preprint
  quant-ph/0402131}\ } (\bibinfo {year} {2004})}\BibitemShut {NoStop}%
\bibitem [{\citenamefont {Cai}\ \emph {et~al.}(2018)\citenamefont {Cai},
  \citenamefont {Guo}, \citenamefont {Lin}, \citenamefont {Zuo},\ and\
  \citenamefont {Yu}}]{CGL+18}%
  \BibitemOpen
  \bibfield  {author} {\bibinfo {author} {\bibfnamefont {B.}~\bibnamefont
  {Cai}}, \bibinfo {author} {\bibfnamefont {G.}~\bibnamefont {Guo}}, \bibinfo
  {author} {\bibfnamefont {S.}~\bibnamefont {Lin}}, \bibinfo {author}
  {\bibfnamefont {H.}~\bibnamefont {Zuo}},\ and\ \bibinfo {author}
  {\bibfnamefont {C.}~\bibnamefont {Yu}},\ }\href@noop {} {\bibfield  {journal}
  {\bibinfo  {journal} {IEEE Photonics Journal}\ }\textbf {\bibinfo {volume}
  {10}},\ \bibinfo {pages} {1} (\bibinfo {year} {2018})}\BibitemShut {NoStop}%
\bibitem [{\citenamefont {He}\ and\ \citenamefont {Ma}(2016)}]{HM16}%
  \BibitemOpen
  \bibfield  {author} {\bibinfo {author} {\bibfnamefont {Y.-F.}\ \bibnamefont
  {He}}\ and\ \bibinfo {author} {\bibfnamefont {W.-P.}\ \bibnamefont {Ma}},\
  }\href@noop {} {\bibfield  {journal} {\bibinfo  {journal} {Quantum
  Information Processing}\ }\textbf {\bibinfo {volume} {15}},\ \bibinfo {pages}
  {5023} (\bibinfo {year} {2016})}\BibitemShut {NoStop}%
\bibitem [{\citenamefont {Huang}\ \emph {et~al.}(2014)\citenamefont {Huang},
  \citenamefont {Wen}, \citenamefont {Liu}, \citenamefont {Gao},\ and\
  \citenamefont {Sun}}]{HWL++14}%
  \BibitemOpen
  \bibfield  {author} {\bibinfo {author} {\bibfnamefont {W.}~\bibnamefont
  {Huang}}, \bibinfo {author} {\bibfnamefont {Q.-Y.}\ \bibnamefont {Wen}},
  \bibinfo {author} {\bibfnamefont {B.}~\bibnamefont {Liu}}, \bibinfo {author}
  {\bibfnamefont {F.}~\bibnamefont {Gao}},\ and\ \bibinfo {author}
  {\bibfnamefont {Y.}~\bibnamefont {Sun}},\ }\href@noop {} {\bibfield
  {journal} {\bibinfo  {journal} {Quantum Information Processing}\ }\textbf
  {\bibinfo {volume} {13}},\ \bibinfo {pages} {649} (\bibinfo {year}
  {2014})}\BibitemShut {NoStop}%
\bibitem [{\citenamefont {Walton}\ \emph {et~al.}(2003)\citenamefont {Walton},
  \citenamefont {Abouraddy}, \citenamefont {Sergienko}, \citenamefont {Saleh},\
  and\ \citenamefont {Teich}}]{WAS+03}%
  \BibitemOpen
  \bibfield  {author} {\bibinfo {author} {\bibfnamefont {Z.~D.}\ \bibnamefont
  {Walton}}, \bibinfo {author} {\bibfnamefont {A.~F.}\ \bibnamefont
  {Abouraddy}}, \bibinfo {author} {\bibfnamefont {A.~V.}\ \bibnamefont
  {Sergienko}}, \bibinfo {author} {\bibfnamefont {B.~E.}\ \bibnamefont
  {Saleh}},\ and\ \bibinfo {author} {\bibfnamefont {M.~C.}\ \bibnamefont
  {Teich}},\ }\href@noop {} {\bibfield  {journal} {\bibinfo  {journal}
  {Physical Review Letters}\ }\textbf {\bibinfo {volume} {91}},\ \bibinfo
  {pages} {087901} (\bibinfo {year} {2003})}\BibitemShut {NoStop}%
\bibitem [{\citenamefont {He}\ and\ \citenamefont {Ma}(2017)}]{HM17}%
  \BibitemOpen
  \bibfield  {author} {\bibinfo {author} {\bibfnamefont {Y.}~\bibnamefont
  {He}}\ and\ \bibinfo {author} {\bibfnamefont {W.}~\bibnamefont {Ma}},\
  }\href@noop {} {\bibfield  {journal} {\bibinfo  {journal} {Modern Physics
  Letters B}\ }\textbf {\bibinfo {volume} {31}},\ \bibinfo {pages} {1750015}
  (\bibinfo {year} {2017})}\BibitemShut {NoStop}%
\bibitem [{\citenamefont {Gao}\ \emph {et~al.}(2018)\citenamefont {Gao},
  \citenamefont {Chen},\ and\ \citenamefont {Qian}}]{GCQ18}%
  \BibitemOpen
  \bibfield  {author} {\bibinfo {author} {\bibfnamefont {H.}~\bibnamefont
  {Gao}}, \bibinfo {author} {\bibfnamefont {X.-G.}\ \bibnamefont {Chen}},\ and\
  \bibinfo {author} {\bibfnamefont {S.-R.}\ \bibnamefont {Qian}},\ }\href@noop
  {} {\bibfield  {journal} {\bibinfo  {journal} {Quantum Information
  Processing}\ }\textbf {\bibinfo {volume} {17}},\ \bibinfo {pages} {140}
  (\bibinfo {year} {2018})}\BibitemShut {NoStop}%
\bibitem [{\citenamefont {Li}\ \emph {et~al.}(2008)\citenamefont {Li},
  \citenamefont {Deng},\ and\ \citenamefont {Zhou}}]{LDZ08}%
  \BibitemOpen
  \bibfield  {author} {\bibinfo {author} {\bibfnamefont {X.-H.}\ \bibnamefont
  {Li}}, \bibinfo {author} {\bibfnamefont {F.-G.}\ \bibnamefont {Deng}},\ and\
  \bibinfo {author} {\bibfnamefont {H.-Y.}\ \bibnamefont {Zhou}},\ }\href@noop
  {} {\bibfield  {journal} {\bibinfo  {journal} {Physical Review A}\ }\textbf
  {\bibinfo {volume} {78}},\ \bibinfo {pages} {022321} (\bibinfo {year}
  {2008})}\BibitemShut {NoStop}%
\bibitem [{\citenamefont {Takeuchi}\ \emph {et~al.}(2016)\citenamefont
  {Takeuchi}, \citenamefont {Fujii}, \citenamefont {Ikuta}, \citenamefont
  {Yamamoto},\ and\ \citenamefont {Imoto}}]{TFI+16}%
  \BibitemOpen
  \bibfield  {author} {\bibinfo {author} {\bibfnamefont {Y.}~\bibnamefont
  {Takeuchi}}, \bibinfo {author} {\bibfnamefont {K.}~\bibnamefont {Fujii}},
  \bibinfo {author} {\bibfnamefont {R.}~\bibnamefont {Ikuta}}, \bibinfo
  {author} {\bibfnamefont {T.}~\bibnamefont {Yamamoto}},\ and\ \bibinfo
  {author} {\bibfnamefont {N.}~\bibnamefont {Imoto}},\ }\href@noop {}
  {\bibfield  {journal} {\bibinfo  {journal} {Physical Review A}\ }\textbf
  {\bibinfo {volume} {93}},\ \bibinfo {pages} {052307} (\bibinfo {year}
  {2016})}\BibitemShut {NoStop}%
\bibitem [{\citenamefont {Wang}\ \emph
  {et~al.}(2022{\natexlab{b}})\citenamefont {Wang}, \citenamefont {Chen},\ and\
  \citenamefont {Sun}}]{WCS22}%
  \BibitemOpen
  \bibfield  {author} {\bibinfo {author} {\bibfnamefont {P.}~\bibnamefont
  {Wang}}, \bibinfo {author} {\bibfnamefont {X.}~\bibnamefont {Chen}},\ and\
  \bibinfo {author} {\bibfnamefont {Z.}~\bibnamefont {Sun}},\ }\href@noop {}
  {\bibfield  {journal} {\bibinfo  {journal} {Physics Letters A}\ }\textbf
  {\bibinfo {volume} {446}},\ \bibinfo {pages} {128291} (\bibinfo {year}
  {2022}{\natexlab{b}})}\BibitemShut {NoStop}%
\bibitem [{\citenamefont {Li}\ \emph {et~al.}(2016)\citenamefont {Li},
  \citenamefont {Wang}, \citenamefont {Zhang}, \citenamefont {Baagyere},
  \citenamefont {Qin}, \citenamefont {Xiong},\ and\ \citenamefont
  {Zhan}}]{LWZ+16}%
  \BibitemOpen
  \bibfield  {author} {\bibinfo {author} {\bibfnamefont {D.-f.}\ \bibnamefont
  {Li}}, \bibinfo {author} {\bibfnamefont {R.-j.}\ \bibnamefont {Wang}},
  \bibinfo {author} {\bibfnamefont {F.-l.}\ \bibnamefont {Zhang}}, \bibinfo
  {author} {\bibfnamefont {E.}~\bibnamefont {Baagyere}}, \bibinfo {author}
  {\bibfnamefont {Z.}~\bibnamefont {Qin}}, \bibinfo {author} {\bibfnamefont
  {H.}~\bibnamefont {Xiong}},\ and\ \bibinfo {author} {\bibfnamefont
  {H.}~\bibnamefont {Zhan}},\ }\href@noop {} {\bibfield  {journal} {\bibinfo
  {journal} {Quantum Information Processing}\ }\textbf {\bibinfo {volume}
  {15}},\ \bibinfo {pages} {4819} (\bibinfo {year} {2016})}\BibitemShut
  {NoStop}%
\bibitem [{\citenamefont {ho~Hong}\ \emph {et~al.}(2017)\citenamefont
  {ho~Hong}, \citenamefont {Heo}, \citenamefont {Jang},\ and\ \citenamefont
  {Kwon}}]{HCJ+17}%
  \BibitemOpen
  \bibfield  {author} {\bibinfo {author} {\bibfnamefont {C.}~\bibnamefont
  {ho~Hong}}, \bibinfo {author} {\bibfnamefont {J.}~\bibnamefont {Heo}},
  \bibinfo {author} {\bibfnamefont {J.~G.}\ \bibnamefont {Jang}},\ and\
  \bibinfo {author} {\bibfnamefont {D.}~\bibnamefont {Kwon}},\ }\href@noop {}
  {\bibfield  {journal} {\bibinfo  {journal} {Quantum Information Processing}\
  }\textbf {\bibinfo {volume} {16}},\ \bibinfo {pages} {236} (\bibinfo {year}
  {2017})}\BibitemShut {NoStop}%
\bibitem [{\citenamefont {Kang}\ \emph {et~al.}(2020)\citenamefont {Kang},
  \citenamefont {Heo}, \citenamefont {Hong}, \citenamefont {Yang},
  \citenamefont {Moon},\ and\ \citenamefont {Han}}]{KHH+20}%
  \BibitemOpen
  \bibfield  {author} {\bibinfo {author} {\bibfnamefont {M.-S.}\ \bibnamefont
  {Kang}}, \bibinfo {author} {\bibfnamefont {J.}~\bibnamefont {Heo}}, \bibinfo
  {author} {\bibfnamefont {C.-H.}\ \bibnamefont {Hong}}, \bibinfo {author}
  {\bibfnamefont {H.-J.}\ \bibnamefont {Yang}}, \bibinfo {author}
  {\bibfnamefont {S.}~\bibnamefont {Moon}},\ and\ \bibinfo {author}
  {\bibfnamefont {S.-W.}\ \bibnamefont {Han}},\ }\href@noop {} {\bibfield
  {journal} {\bibinfo  {journal} {Quantum Information Processing}\ }\textbf
  {\bibinfo {volume} {19}},\ \bibinfo {pages} {24} (\bibinfo {year}
  {2020})}\BibitemShut {NoStop}%
\bibitem [{\citenamefont {Jiang}\ \emph {et~al.}(2021)\citenamefont {Jiang},
  \citenamefont {Zhou},\ and\ \citenamefont {Hu}}]{JZH21}%
  \BibitemOpen
  \bibfield  {author} {\bibinfo {author} {\bibfnamefont {S.}~\bibnamefont
  {Jiang}}, \bibinfo {author} {\bibfnamefont {R.-G.}\ \bibnamefont {Zhou}},\
  and\ \bibinfo {author} {\bibfnamefont {W.}~\bibnamefont {Hu}},\ }\href@noop
  {} {\bibfield  {journal} {\bibinfo  {journal} {International Journal of
  Theoretical Physics}\ }\textbf {\bibinfo {volume} {60}},\ \bibinfo {pages}
  {3353} (\bibinfo {year} {2021})}\BibitemShut {NoStop}%
\bibitem [{\citenamefont {Wu}\ \emph {et~al.}(2021)\citenamefont {Wu},
  \citenamefont {Chang}, \citenamefont {Guo},\ and\ \citenamefont
  {Lin}}]{WCG+21}%
  \BibitemOpen
  \bibfield  {author} {\bibinfo {author} {\bibfnamefont {Y.-T.}\ \bibnamefont
  {Wu}}, \bibinfo {author} {\bibfnamefont {H.}~\bibnamefont {Chang}}, \bibinfo
  {author} {\bibfnamefont {G.-D.}\ \bibnamefont {Guo}},\ and\ \bibinfo {author}
  {\bibfnamefont {S.}~\bibnamefont {Lin}},\ }\href@noop {} {\bibfield
  {journal} {\bibinfo  {journal} {International Journal of Theoretical
  Physics}\ ,\ \bibinfo {pages} {4066}} (\bibinfo {year} {2021})}\BibitemShut
  {NoStop}%
\bibitem [{\citenamefont {Liu}\ \emph {et~al.}(2024)\citenamefont {Liu},
  \citenamefont {Luo}, \citenamefont {Yu}, \citenamefont {Wang}, \citenamefont
  {Wang}, \citenamefont {Hu}, \citenamefont {Li}, \citenamefont {Zheng},
  \citenamefont {Yao}, \citenamefont {Yan} \emph {et~al.}}]{LLY+24}%
  \BibitemOpen
  \bibfield  {author} {\bibinfo {author} {\bibfnamefont {J.-L.}\ \bibnamefont
  {Liu}}, \bibinfo {author} {\bibfnamefont {X.-Y.}\ \bibnamefont {Luo}},
  \bibinfo {author} {\bibfnamefont {Y.}~\bibnamefont {Yu}}, \bibinfo {author}
  {\bibfnamefont {C.-Y.}\ \bibnamefont {Wang}}, \bibinfo {author}
  {\bibfnamefont {B.}~\bibnamefont {Wang}}, \bibinfo {author} {\bibfnamefont
  {Y.}~\bibnamefont {Hu}}, \bibinfo {author} {\bibfnamefont {J.}~\bibnamefont
  {Li}}, \bibinfo {author} {\bibfnamefont {M.-Y.}\ \bibnamefont {Zheng}},
  \bibinfo {author} {\bibfnamefont {B.}~\bibnamefont {Yao}}, \bibinfo {author}
  {\bibfnamefont {Z.}~\bibnamefont {Yan}}, \emph {et~al.},\ }\href@noop {}
  {\bibfield  {journal} {\bibinfo  {journal} {Nature}\ }\textbf {\bibinfo
  {volume} {629}},\ \bibinfo {pages} {579} (\bibinfo {year}
  {2024})}\BibitemShut {NoStop}%
\end{thebibliography}%

\end{document}